\documentclass[prb,twocolumn,showpacs,preprintnumbers]{revtex4-2}

\usepackage[dvipdfmx]{graphicx}
\usepackage{amsmath,amssymb}
\usepackage{times}
\usepackage{physics}
\usepackage{xcolor}
\usepackage{colortbl}
\usepackage{array, tabularx, diagbox}
\usepackage{ulem}
\usepackage{multirow}
\usepackage{bm}
\usepackage{xr}
\usepackage{comment}

\usepackage[colorlinks=true,urlcolor=blue,citecolor=blue,linkcolor=blue,breaklinks=true]{hyperref}
\usepackage{cleveref}

\usepackage{titlesec}

\newcommand{\textr}[1]{{\textcolor{red}{#1}}}
\newcommand{\textb}[1]{{\textcolor{cyan}{#1}}}
\definecolor{caribbeangreen}{rgb}{0.0, 0.8, 0.6}

\definecolor{mygreen}{rgb}{0.0, 0.6, 0.1}

\newcolumntype{M}[1]{>{\centering\arraybackslash}p{#1}}

\newcommand{\expect}[1]{\langle#1\rangle}
\newcommand{\bme}{\bm{e}}

\newcommand{\gbar}{\overline{G}}
\newcommand{\muFM}{\mu_{\textrm{FM}}}
\newcommand{\muDLM}{\mu_{\textrm{DLM}}}
\newcommand{\muB}{\mu_{\textrm{B}}}
\newcommand{\PES}{{K$_4$Al$_3$(SiO$_4$)$_3$}}

\newcommand{\CoTS}{{Co$_{1/3}$}Ta{S$_2$}}
\newcommand{\NiTS}{{Ni$_{1/3}$}Ta{S$_2$}}
\newcommand{\TMTS}{{TM$_{1/3}$}Ta{S$_2$}}
\newcommand{\Jab}{J_{ab}}
\newcommand{\Jc}{J_{c}}

\begin{document}

\title{Calculation of the biquadratic spin interactions based on the spin cluster expansion for \textit{ab initio} tight-binding models}
\author{Tatsuto Hatanaka$^{1,2}$}
\email{hatanaka-tatsuto346@g.ecc.u-tokyo.ac.jp}
\author{Juba Bouaziz$^{3}$}
\email{jbouaziz@g.ecc.u-tokyo.ac.jp}
\author{Takuya Nomoto$^{4}$}
\email{tnomoto@tmu.ac.jp}
\author{Ryotaro Arita$^{2,3}$}
\email{arita@riken.jp}
\affiliation{$^1$ Department of Applied Physics, The University of Tokyo, Bunkyo-ku, Tokyo, 113-8656, Japan \\
$^2$ RIKEN Center for Emergent Matter Science (CEMS), Wako 351-0198, Japan\\
$^3$ Department of Physics, The University of Tokyo, Bunkyo-ku, Tokyo, 113-0033, Japan\\
$^4$ Department of Physics, Tokyo Metropolitan University, Hachioji, Tokyo 192-0397, Japan}
\date{\today}

\begin{abstract}
We develop a calculation scheme using \textit{ab initio} tight-binding Hamiltonians to evaluate biquadratic magnetic interactions.
This approach relies on the spin cluster expansion combined with the disordered local moment (DLM) method, originally developed within the multiple scattering Korringa-Kohn-Rostoker method.
Applying it to a single-orbital Hubbard model with two sublattices, we show that the evaluated DLM biquadratic interactions are in good agreement with those obtained from the strongly correlated limit, demonstrating the wide applicability of the method to various magnetic systems with large local moments.
We then apply it to the \textit{ab initio} tight-binding models for elemental magnetic metals; the resulting magnetic interactions align well with previous literature.
Finally, we explore its performance in more complex compounds, such as transition metal dichalcogenides with intercalation of 3\textit{d} transition metals and potassium electrosodalite.
The obtained results for both compounds show good agreement with experiments.
The present approach offers a convenient \textit{ab initio} path for evaluating biquadratic interactions and understanding the electronic mechanisms controlling them.
\end{abstract}

\maketitle

%------------------------------------------------------------------------
%------------------------------------------------------------------------
%------------------------------------------------------------------------
%------------------------------------------------------------------------
\section{Introduction}\label{sec:intro}

Recently, antiferromagnetic materials with nontrivial spin arrangements have attracted broad interest due to their unique properties, making them suitable for technological applications~\cite{afm-spintronics-1, afm-spintronics-2, afm-spintronics-3}.
These complex spin configurations often arise from spin interactions beyond what the standard Heisenberg model accounts for~\cite{Heisenberg1926-nf}.
For instance, the Dzyaloshinskii-Moriya (DM) interaction~\cite{dzyaloshinsky, moriya}, which originates from the spin-orbit coupling and inversion symmetry breaking, is well-known for inducing noncollinear chiral spin structures such as magnetic skyrmions~\cite{Bogdanov1989,nagaosa2013}.
In addition, the biquadratic interaction~\cite{Rodbell1963-jk, Huang1964-at, Harris1963-wb}, a direct extension of the bilinear Heisenberg interaction, is also essential in stabilizing such complex spin 
arrangements~\cite{Hayami2021, Simon2020}. The biquadratic interaction can drive the system to a non-collinear configuration even in the absence of the spin-orbit coupling, its effects on various spin systems have been extensively investigated from both theoretical and experimental perspectives~\cite{Bhatt1998-kj, eff-blbq-hym, Okumura2022-ct, Seo2021-vy,ucu5-bq, paddison2023cubic, cts-seki, cts-ppark}.

In order to quantify the strength of the biquadratic interaction within a first principles framework, it is highly desirable to derive parameters of the spin Hamiltonian from realistic \textit{ab initio} methods~\cite{Szilva2023-lr}.
For challenging tasks, the infinitesimal rotation method or Liechtenstein-Katsnelson-Antropov-Gubanov (LKAG) method~\cite{oth, lkag, local-force-nmt} is established as a commonly used approach to extract spin Hamiltonian parameters and has been successfully applied to several material classes, including $3d$-transition metal, $4f$-elements, and transition metal oxides~\cite{Szilva2023-lr, sakuma-3d, sakuma-alloy, nickelates, gd-skx, htnk, Bouaziz2022, Bouaziz2024, Miao2024, htnk, Logemann2017}. 
The LKAG approach consists of mapping the energy variation of magnetic quantum systems due to an infinitesimal rotation onto a classical system with localized spins~\cite{oth}.
The relativistic extension of the LKAG approach gives access to the DM interaction and two-ion anisotropy~\cite{Udvardi2003,Ebert2009}. The asymptotic behavior of the obtained pair exchange interactions obtained by the LKAG approach is well-understood for both the strongly correlated and itinerant limits\cite{Solovyev1999-xj}.

The LKAG approach has nonetheless a few shortcomings:
First, the band energy variation due to an infinitesimal rotation is mapped onto bilinear interactions, incorporating contributions from spin interactions of a higher-order nature. Second, the method is based on a magnetically ordered reference state, which may lead to different results depending on the chosen reference state~\cite{Lounis2010, Szilva2013}; hence, this approach is suitable for a local mapping~\cite{Dias2021}.
We note that while extended methods have been proposed to compute higher-order or multi-spin interactions~\cite{szilva, extension}, the reference state dependence issue remains present.

An alternative approach to determine the exchange parameters of the classical spin model consists of fitting the total energy dependence of multiple spin configurations~\cite{Kurz2001,grytsiuk2020}.
This method enables the self-consistent evaluation of arbitrary spin configurations through the application of magnetic constraining fields~\cite{ujfalussy1999} and derives exchange parameters to arbitrary orders.
The drawbacks of this technique are the necessity for large multiple-atom supercells to handle complex spin configurations, the high computational cost, and the accuracy of the fitting procedure, which may lead to spurious results when dealing with small magnetic interactions in the meV range.

A computationally efficient method that accesses the pair and higher-order magnetic interactions without supercell limitations and is almost independent of the reference state is highly desirable.
The relativistic disordered local moment (RDLM)~\cite{dlm, Staunton2006, patrick2018, Bouaziz2023}, combined with the spin cluster expansion (SCE)~\cite{sce1,sce2}, provides an SCE-RDLM scheme for a systematic evaluation of magnetic interactions of arbitrary order~\cite{sce-rdlm, sce-rdlm-eg1, sce-rdlm-eg2, sce-rdlm-eg3, sce-rdlm-eg4, Deak2011-jl, Staunton2014}.
In the SCE approach, the energy surface of a classical spin system with frozen length is systematically expanded into spin clusters using a real spherical harmonics basis.
The basis describes the magnetic orientations of the spin moments, and the expansion coefficients encode the magnetic interactions. The order of the basis function dictates the form of the spin Hamiltonian~\cite{sce1}.
By combining this with the DLM picture, in which the paramagnetic state is considered, we can systematically calculate arbitrary-order and multi-site spin interactions with less computational cost than the fitting approach.
The SCE-RDLM formalism was originally developed for the multiple scattering theory based on the Green’s function methods, such as the Korringa-Kohn-Rostoker (KKR) method~\cite{zeller2014}, and implemented in the spin density functional theory (SDFT).
Additionally, the dependence on the reference state has been completely eliminated in the framework of the KKR method.

In this work, we formulate SCE-RDLM for \textit{ab initio} tight-binding models.
Since our approach only requires tight-binding parameters, it is compatible with a variety of \textit{ab initio} frameworks employing different basis functions, such as plane-wave and pseudo-atomic localized basis functions~\cite{Ozaki2003-bo, Ozaki2004-yz}.
Although our tight-binding parameters are obtained from a specific magnetic order and, therefore, carry some reference state dependence, the electronic Green functions used in the determination of the spin interaction parameters are obtained via the DLM method.
Consequently, our formalism avoids the reference state dependence except during the construction of the tight-binding model from first principles, offering a significant advantage over the LKAG method.
Therefore, the presented approach is a useful tool to understand and predict the physical properties of magnetic materials with complex spin configurations that arise from higher-order spin interactions.

The paper is structured as follows:
In Section \ref{sec:formulation}, we formulate SCE-DLM, the nonrelativistic version of SCE-RDLM, for \textit{ab initio} tight-binding models.
The presented approach is applicable to arbitrary spin interactions.
Our primary emphasis is on the biquadratic exchange interaction, which plays a central role in stabilizing non-coplanar order in antiferromagnets without the assistance of relativistic effects (DM interaction).
This section also discusses the reference state dependence of our approach, which is completely eliminated in the original method with KKR.
The computational details and setups are present in Section \ref{sec:comp-detail}.
Section \ref{sec:benchmark} is dedicated to the benchmarking, validation, and discussions of our approach: we first apply it to the one-dimensional Hubbard model with two sublattices.
We analyze the asymptotic behavior of the bilinear and biquadratic interactions calculated for the limit of strong correlation and confirm that the interactions evaluated by SCE-DLM align with those evaluated for the effective quantum spin model.
Next, we apply SCE-DLM to three magnets using the \textit{ab initio} tight-binding approach.
The first is bcc Fe, where local magnetic moments are well-defined.
We compare our results with previous theoretical ones on the biquadratic interaction for this magnet.
The second and third are more itinerant magnets, where the local moment picture is less valid, i.e., hcp and fcc Co.
We compare our results with the ones obtained using the KKR method for the above magnets and discuss the dependence on the starting magnetic order.
In Section \ref{sec:compounds}, we demonstrate the applicability of our method to more complex compounds.
The first example is Co$_{1/3}$TaS$_2$, a transition metal dichalcogenide (TMD) intercalated with 3$d$ transition metals.
In this compound, a non-coplanar spin structure has been experimentally reported by multiple groups\textb{~\cite{Takagi2023-uw, Park2023-ys}}.
The biquadratic interaction has been proposed as the driving mechanism behind this magnetic configuration.
Another example of a complex compound is potassium electrosodalite, i.e., K$_4$Al$_3$(SiO$_4$)$_3$, which highlights the versatility of our method.
Specifically, this example demonstrates that our method is also applicable to systems in which magnetic moments are polarized in interstitial regions rather than on atomic sites.
Finally, in Section \ref{sec:conclusion}, we conclude and summarize our work.

%------------------------------------------------------------------------
%------------------------------------------------------------------------
%------------------------------------------------------------------------
%------------------------------------------------------------------------
\section{Formulation}\label{sec:formulation}
%------------------------------------------------------------------------
%------------------------------------------------------------------------
\subsection{Spin Cluster Expansion}\label{subsec:sce}
The spin cluster expansion (SCE) developed by Drautz and F\"{a}hnle \cite{sce1,sce2} provides a tool to expand systematically the energy of a many-body classical spin system by introducing clusters consisting of several spins.
The SCE employs a complete orthonormal basis set, namely the real spherical harmonics $Y_{L=(l,m)}(\bm{e})$ for a unit vector $\bm{e}$ of a classical spin.
The orthogonality relations are defined as follows~\cite{real-sph-harm}:
\begin{align}
    \frac{1}{4\pi}\int\dd^2{\bme}~Y_{L}(\bme)Y_{L'}(\bme) = \delta_{L,L'}
\end{align}
where $\int\dd^2\bm{e}$ stands for integration over the surface of a unit sphere.

The basis functions for a cluster $C$ are built from a multiplication of the basis functions of every single spin, resulting in:
\begin{align}\label{eq:cluster-basis}
    \Phi^{\{L\}}_{C}(\{\bm{e}\})&=\frac{1}{(4\pi)^{(N-n)/2}}\prod_{i\in{C}}Y_{L_{i}}(\bm{e}_{i})
\end{align}
where $N$ refers to the number of all spins in the system, $n$ indicates the size of the cluster, $\{\bm{e}\}$ is an array that represents the spin directions, and $\{L\}$ is an array containing the combined orbital index $L=(l,m)$ of each spin in the cluster $C$.
These basis functions of the clusters also form a complete and orthonormal basis. The grand potential of the spin system can then be expanded in terms of the cluster basis as:
\begin{align}
    \Omega(\{\bm{e}\}) =&\,\Omega_0 + \sum_{C}\sum_{\{L\}}J^{\{L\}}_{C}\Phi^{\{L\}}_{C}(\{\bm{e}\}) \label{eq:pes-expand}
\end{align}    
where the expansion coefficients of each cluster are defined as:
\begin{align}
    J^{\{L\}}_{C} =& \bra{\Phi^{\{L\}}_{C}}\ket{\Omega}\label{eq:inner-prod-coeff}\\
    \bra{f}\ket{g} = \idotsint&\qty[\prod_{i=1}^N\qty(\dd^2\bm{e}_i)] f(\{\bm{e}\})g(\{\bm{e}\})\label{eq:def-inner-prod}
\end{align}
where $\Omega_0$ stands for a constant offset independent of the spin configuration.
In Eq.~(\ref{eq:inner-prod-coeff}), we use the useful Dirac bra-ket notation, and the inner product in this notation is defined in Eq.~(\ref{eq:def-inner-prod}) for functions $f$ and $g$ which depend on the spin configuration $\{\bm{e}\}$.

Our focus lies in two site interactions (pair and biquadratic). Therefore, we consider the grand potential expanded up to the two-spin clusters. We note that the expansion for clusters consisting of multiple sites can be performed in a similar way:
\begin{align}
    \Omega(\{\bm{e}\}) =&\,\Omega_0+\sum_{i}\sum_{L\neq(0,0)}J_i^L Y_{L}(\bm{e}_i)\nonumber\\
    &+\frac{1}{2}\sum_{i\neq j}\sum_{L,L'\neq(0,0)}J_{ij}^{LL'} Y_{L}(\bm{e}_i)Y_{L'}(\bm{e}_j)
\end{align}
Each expansion coefficient for one/two spin clusters can be evaluated %following 
using 
the inner product in Eqs.~(\ref{eq:inner-prod-coeff}) and (\ref{eq:def-inner-prod}).
\begin{align}
    J_i^L &= \int \dd^2 \bm{e}_i \expect{\Omega}_{\bme_i}Y_L(\bme_i)\label{eq:int-1}\\
    J_{ij}^{LL'} &= \iint \dd^2 \bme_i \dd^2 \bme_j\expect{\Omega}_{\bme_i\bme_j}Y_{L}(\bme_i)Y_{L'}(\bme_j)\label{eq:int-2}\\
    \expect{\Omega}_{C} &= \frac{1}{(4\pi)^{N-n}}\idotsint\qty[\prod_{i\notin{C}}\qty(\dd^2\bme_i)]\Omega(\{\bm{e}\})\label{eq:int-c}
\end{align}
where $\expect{\Omega}_{C}$ stands for the expectation value calculated by integrating all solid angles except for spins in the cluster $C$.
The directions of the spins outside the cluster are integrated over, and the evaluated coefficients are independent of a reference state.
Since we can not straightforwardly calculate Eq.~(\ref{eq:int-c}) for many-body systems, we use the disordered local moment state, which is discussed in Sec.~\ref{subsec:dlm}.

%------------------------------------------------------------------------
%------------------------------------------------------------------------
\subsection{Disordered Local Moment}\label{subsec:dlm}
The coherent potential approximation (CPA) was initially introduced to compute the electronic structure of disordered alloys~\cite{Soven1967}.
The CPA relies on embedding impurities described by a single site scattering matrix into an effective medium and imposing that the additional impurity scattering should vanish. The disordered local moment (DLM) approach employs a similar approach to CPA to incorporate the effect of transverse spin fluctuation~\cite{dlm}.
The DLM relies on the approximation that the local moments $\{\bm{e}\}$ motion is much slower than the electronic ones. The CPA averaging is done on the magnetic moment orientations and models an electronic structure in the paramagnetic state with randomly oriented spins.
While the DLM method was originally developed for the KKR method, it can similarly be formulated for the tight-binding Hamiltonians~\cite{wannier-cpa}.

Here, we consider the multi-orbital tight-binding Hamiltonian defined as:
\begin{align}\label{eq:tb-ham}
    \mathcal{H} = \sum_{i\ell'\sigma,jm'\sigma'}\qty(H_{i\ell'\sigma,jm'\sigma'}\hat{c}^{\dagger}_{i\ell'\sigma}\hat{c}_{jm'\sigma'}+\mathrm{h.c.}),
\end{align}
where $(i, j), (\ell', m')$, and $(\sigma,\sigma')$ are the indices of sites, orbitals, and spins, respectively.
The operator $\hat{c}_{i\ell'\sigma}/\hat{c}^{\dagger}_{i\ell'\sigma}$ stands for the annihilation/creation operator of an electron specified with the degrees of freedom $(i\ell'\sigma)$.
We divide each component of the Hamiltonian into the spin-independent off-site hopping $t$ and the on-site magnetic potential term $v$,
\begin{align}\label{eq:tb-component}
    H_{i\ell'\sigma,jm'\sigma'} = t_{i\ell',jm'}\delta_{\sigma,\sigma'} + \delta_{ij}v^{i}_{\ell',m'}[\bme_{i}\cdot\bm{\sigma}]_{\sigma\sigma'},
\end{align}
where $\bme_{i}, \bm{\sigma}$ is the direction of a spin at the site $i$ and the Pauli matrix vector, respectively.
Here, we assume that there are no spin-dependent hopping terms and that the spin-dependent potentials $v$'s are a local quantity.
Although there could be non-local spin-dependent potentials in the tight-binding Hamiltonian constructed from first principles, the latter are not included to 
improve computational efficiency~\cite{local-force-nmt}.
We denote the former term of Eq.~(\ref{eq:tb-component}) as $H_0$ and the latter as $V$ such that $H=H_0+V$.

In the DLM method, we consider the virtual state with randomly oriented spins and introduce the self-energy $\Sigma$, instead of the spin-dependent potential $V$, corresponding to the effective potential of such a disordered state as follows:
\begin{align}
    H_c &= H_0 + \Sigma\\
    H &= H_c + (V-\Sigma),
\end{align}
where $H_c$ indicates the Hamiltonian of the DLM state.
Note that the introduced self-energy $\Sigma$ is a local and spin-independent complex quantity so that $\Sigma_{i\ell'\sigma,jm'\sigma'} = \delta_{ij}{\delta_{\sigma\sigma^\prime}}\tilde{\Sigma}^i_{\ell'\sigma,m'\sigma'}$.
The Green's functions in the real space are given as follows:
\begin{align}
    \gbar(\epsilon) &= \qty[\epsilon-H_c]^{-1}\\
    G(\epsilon) &= \qty[\epsilon-H]^{-1}\nonumber\\
    &= \gbar\qty[1+(V-\Sigma)\gbar]^{-1},
\end{align}
We also introduce the scattering matrix $T$ as follows:
\begin{align}
    T_{i}(\bme_i) = \qty(V_i(\bm{e}_i)-\tilde{\Sigma}^i) \qty[1-\gbar_{ii}\qty[V_i(\bm{e}_i)-\tilde{\Sigma}^i]]^{-1}\label{eq:def-t-matrix}
\end{align}
where $V_i(\bme_i)$ stands for the magnetic potential of the spin at site $i$ with the orientation $\bme_i$, namely $[V_i(\bme_i)]_{\ell'\sigma,m'\sigma'}=v^{i}_{\ell',m'}[\bme_{i}\cdot\bm{\sigma}]_{\sigma\sigma'}$, and the scattering matrix $T_{i}(\bme_i)$ has the same degrees of freedom with those of $V_i(\bme_i)$.
We then can formulate the CPA condition for the tight-binding Hamiltonian with the single-site approximation~\cite{wannier-cpa},
\begin{align}
    \frac{1}{4\pi}\int \dd^2\bme_{i}T_i(\bme_i)=0 \label{eq:cpa-condition}.
\end{align}
In the absence of the relativistic spin-orbit coupling (SOC), the CPA condition in Eq.(\ref{eq:cpa-condition}) is expressed as follows (Ising DLM):
\begin{align}
    \frac{T_i(\hat{\bm{z}}) + T_i(-\hat{\bm{z}})}{2} = 0,
\end{align}
where $\hat{\bm{z}}$ is a unit vector along the $z$-axis.

In practice, we determine the self-energy and the chemical potential in a self-consistent manner~\cite{wannier-cpa}. 
The chemical potential $\mu_c$ of the DLM state is set by the conservation condition for the number of electrons below,
\begin{align}
    N = -\frac{1}{\pi}&\int\dd\epsilon f(\epsilon)\Im\Tr\gbar(\epsilon)\label{eq:determine-mu}\\
    f(\epsilon) &= \frac{1}{1+e^{\beta(\epsilon-\mu_c)}}
\end{align}
where $\Tr$ stands for taking a trace over all indices of the sites, orbitals, and spins, and $f(\epsilon)$ is the Fermi distribution function.

Let us now clarify the dependence on the reference state in our approach.
When constructing the tight-binding model from first principles, the local magnetic potentials $v_i$ are fixed.
Unlike the KKR method, where the DLM state is obtained self-consistently, including such local exchange splittings, the reference state dependence in our approach partially remains due to these fixed magnetic potentials.
However, once the tight-binding parameters in Eq. (\ref{eq:tb-ham}) are obtained, the DLM state is determined in a self-consistent manner, as described above.
Therefore, this ensures that no further reference state dependence is introduced during this step and subsequent calculations of spin interactions.

%------------------------------------------------------------------------
%------------------------------------------------------------------------
\subsection{SCE-DLM Scheme}\label{subsec:sce-dlm}
The restricted averages of the grand potential $\expect{\Omega}_{\bme_i\bme_j}$ are the central quantities of interest to determine the two-site magnetic interactions~\cite{sce-rdlm}.
$\expect{\Omega}_{\bme_i\bme_j} $ are obtained using the Lloyd's formula~\cite{lloyd1, lloyd2} for the reference state, i.e., the DLM state. Within our tight-binding formulation, we obtain the following expression:
\begin{align}
    \expect{\Omega}_{\bme_i\bme_j} =&\, \Omega_0 -\frac{1}{\pi}\Im\int\dd\epsilon{f}(\epsilon)\Biggl[\ln\det\qty[1+T_i(\bme_i)\gbar_{ii}]\nonumber\\
    &\,+\ln\det\qty[1+T_j(\bme_j)\gbar_{jj}]\nonumber\\
    &\,+\sum_{n\neq i,j}\int\dd^2\bme_n\ln\det\qty[1+T_n(\bme_n)\gbar_{nn}]\nonumber\\
    &\,+\sum_{k=1}^{\infty}\frac{1}{k}\Tr[\left(T_{i}(\bm{e}_i)\gbar_{ij}T_{j}(\bm{e}_j)\gbar_{ji}\right)^k]\Biggl]
\end{align}
where $\Omega_0$ stands for the energy of the DLM state.
The second and third terms indicate the corrections in the integrated density of states from the DLM state due to fixing the orientations of spins in the cluster consisting of spin $i$ and $j$ corresponding to the orientations $\bm{e}_i$ and $\bm{e}_j$.
The fourth term takes into account the contributions of spins outside the cluster, integrated over spin directions via Eq. (\ref{eq:int-c}).
The fifth term denotes the contributions from the scatterings within the two-spin cluster.
The two site expansion coefficients are then obtained using Eq.~(\ref{eq:int-2}), the integration with the spherical harmonics yields the expansion coefficients for the two-spin clusters~\cite{sce-rdlm},
\begin{align}
    J^{LL'}_{ij}=&\,-\frac{1}{\pi}\Im \int\dd\epsilon{f}(\epsilon) \iint\dd^2\bme_i\dd^2\bme_j Y_L(\bme_i)Y_{L'}(\bme_j)\nonumber\\
    &\,\times\ln\det\qty[1-T_{i}(\bm{e}_i)\gbar_{ij}T_{j}(\bm{e}_j)\gbar_{ji}].\label{eq:jijl}
\end{align}

In order to map the energy expanded with the one/two-spin clusters in Eq.~(\ref{eq:pes-expand}) to the following classical spin Hamiltonian, 
\begin{align}
    \mathcal{H} = -2\sum_{\langle{i,j}\rangle}\qty[J_{ij}(\bme_i\cdot\bme_j) + B_{ij}(\bme_i\cdot\bme_j)^2]\label{eq:classical-ham},
\end{align}
we use the sum rule for the spherical harmonics below,
\begin{align}
    \frac{4\pi}{2l+1}\sum_{m} Y^m_{l}(\bm{e}_{i})Y^m_{l}(\bm{e}_j) = P_{l}(\bm{e}_{i}\cdot\bm{e}_j)
\end{align}
where $P_{l}(x)$ is the Legendre polynomial.
Here, we consider an SOC-free case, so that the Hamiltonian has the SU(2) symmetry. Hence, the expansion coefficients $J_{ij}^{LL'}$ do not depend on $m$, i.e., the spin interactions are isotropic. 
Given that the bilinear (BL) and biquadratic (BQ) interactions in the Hamiltonian in Eq.~(\ref{eq:classical-ham}) correspond to $l=1$ and 2, respectively, we can obtain these parameters from the expansion coefficients $J^{LL'}_{ij}$ as follows:
\begin{align}
    J_{ij}&=\frac{1}{8\pi}\sum_{m=-1}^{1}J_{ij}^{(1,m)(1,m)} =\frac{3}{8\pi}J_{ij}^{(1,0)(1,0)}\label{eq:jijl2jij-1}\\
    B_{ij}&=\frac{3}{16\pi}\sum_{m=-2}^{2}J_{ij}^{(2,m)(2,m)} = \frac{15}{16\pi}J_{ij}^{(2,0)(2,0)}\label{eq:jijl2jij-2}
\end{align}

%------------------------------------------------------------------------
%------------------------------------------------------------------------
%------------------------------------------------------------------------
%------------------------------------------------------------------------
\section{Computational Details}\label{sec:comp-detail}
%------------------------------------------------------------------------
%------------------------------------------------------------------------
\subsection{Benchmarks}\label{subsec:comp-benchmark}
We here summarize the computational details for calculations in Section \ref{sec:benchmark}.

\subsubsection{SCE-DLM scheme}\label{subsubsec:comp-sce-dlm}
In the calculation, the inverse temperature $\beta$ was set to 500 $\textrm{eV}^{-1}$.
To evaluate the Green's function in the reciprocal space, we use 512$\times$1$\times$1 $k$-point grid for the one-dimensional Hubbard model.
For bcc Fe, hcp Co, and fcc Co, we use 24$\times$24$\times$24 $k$-point grid for the primitive cell of each material.
We employ the efficient Lebedev quadrature scheme~\cite{lebedev} in the integration over solid angles. The integration over real energies in Eqs.~(\ref{eq:determine-mu}) and (\ref{eq:jijl}) can be transformed to the summation over the fermionic Matsubara poles by analytical continuation.
We use the intermediate representation of the Green's function~\cite{irbasis1,irbasis2} to reduce the computational cost.

%------------------------------------------------------------------------
\subsubsection{Construction of Wannier-based Tight-binding Model}\label{subsubsec:comp-wannier}

We performed SDFT calculations for Fe and Co with the QUANTUM ESPRESSO package~\cite{qe1,qe2} with non-relativistic pseudopotentials in PSlibrary~\cite{pslibrary}.
We used the projector augmented wave method~\cite{paw1,paw2} and the Perdew-Burke-Ernzerhof (PBE) exchange-correlation functional~\cite{pbe}.
The energy cut-off for the plane-wave basis was set to 50 Ry for bcc Fe and 80 Ry for hcp and fcc Co, and a 16$\times$16$\times$16 $k$-point grid was used for the primitive cell.
We set the lattice constant as the experimental value of a = 2.866 \AA~ for bcc Fe, a = 3.550 \AA~ for fcc Co, and a (c) = 2.506 (4.069) \AA~ for hcp Co.

The Wannier functions were constructed using the Wannier90 code~\cite{mlwf1,mlwf2,wannier90-1,wannier90-2}.
The inner window to reproduce the low energy band dispersion of the DFT calculations was set from $E_{\mathrm{F}}-10$ to $E_{\mathrm{F}}+10$ eV for bcc Fe, and to $E_{\mathrm{F}}+3$ for hcp and fcc Co, with $E_{\mathrm{F}}$ being the Fermi energy.
We constructed a nine-orbital model containing one 4$s$, five 3$d$, and three 4$p$ atomic orbitals per atom.
In constructing the Wannier functions, 8$\times$8$\times$8 sampling $k$-point grid was used. The present calculation employs the plane-wave basis, and it should be noted that the construction of the tight-binding model does not depend on the choice of the basis functions of the SDFT calculation.

%------------------------------------------------------------------------
\subsubsection{KKR Method}\label{subsubsec:comp-kkr}
In order to benchmark our \textit{ab initio} tight-binding for bcc Fe, hcp Co, and fcc Co, we performed SDFT calculations using the all-electron full-potential KKR Green function method using the jukkr code~\cite{juKKR:22}, in the scalar relativistic approximation and neglecting spin-orbit coupling.
{The cutoff of the orbital momentum expansion for the KKR Green function is set at $l_{\textrm{max}}=3$.} The self-consistent calculations are 
performed using an energy contour including $51$ energy points and a Fermi-Dirac
smearing of $500$ K with a k-mesh density of 30$\times$30$\times$30. The exchange-correlation is included in the Generalized Gradient approximation (GGA)~\cite{Perdew1992}. 
The paramagnetic (DLM) state is obtained using CPA approximation with $50\%$ spin-up and down magnetic configurations, including charge self-consistency.
The magnetic interactions are computed from the DLM reference state.
The pair interactions are obtained using the infinitesimal rotation method~\cite{lkag}.
The BQ interactions are computed via the fourth-order extension of the infinitesimal rotation method~\cite{extension} with the electronic Green's function of the DLM state.
The pair and BQ interactions are computed on an energy contour, which includes $128$ energy points and a Fermi-Dirac smearing of $200$ K and a 40$\times$40$\times$40 k-mesh.

%------------------------------------------------------------------------
%------------------------------------------------------------------------
\subsection{Application to Complex Compounds}\label{subsec:comp-compounds}
In this subsection, we summarize computational details of calculations in Section \ref{sec:compounds}.

%------------------------------------------------------------------------
\subsubsection{SCE-DLM Scheme}\label{subsubsec:comp-sce-dlm2}
In the calculation, the inverse temperature $\beta$ was set to 600 $\textrm{eV}^{-1}$ for \CoTS~and 500 $\textrm{eV}^{-1}$ for \PES.
It should be noted that 600 $\textrm{eV}^{-1}$ is lower than the experimentally reported temperature of the phase transition to the non-coplanar spin structure in \CoTS.
To evaluate the Green's function in the reciprocal space, we used 20$\times$20$\times$8 and 16$\times$16$\times$16 $k$-point grids for \CoTS and \PES, respectively.
As in the benchmark calculations, we also employed the Lebedev quadrature scheme and the intermediate representation of the Green's function.

%------------------------------------------------------------------------
\subsubsection{Construction of Wannier-based Tight-binding Model}\label{subsubsec:comp-wannier2}

We used the Vienna Ab initio Simulation Package (VASP) code~\cite{Kresse1996-kg} for SDFT calculations of compounds in Section \ref{sec:compounds}.
We employed the PBE exchange-correlation functional proposed and pseudopotentials generated from the projector augmented wave method, and neglected the spin-orbit coupling.
As in the calculations for benchmarks, we also employed the Wannier90 code to construct \textit{ab initio} tight-binding models.

For \CoTS, we set the lattice constants as the experimental value of $a = b = 5.74$ \AA~ and $c = 11.932$ \AA~\cite{Parkin1983-ge}.
The energy cut-off for the plane-wave basis set was set to 600 eV, and we used a 16$\times$16$\times$8 $k$-point grid for the primitive cell.
In the pseudopotentials, the valence electron configurations were 3$p^6$3$d^8$4$s^1$ (Co), 5$p^6$6$d^4$6$s^1$ (Ta), and 3s$^2$3$p^4$ (S), respectively.
We also performed SDFT$+U$ calculations and determined the value of $U = 0.5$ eV to match the magnitude of the magnetic moment of Co with the experimental result~\cite{Takagi2023-uw}.
Details and results of the SDFT$+U$ calculations for \CoTS are given in Appendix~\ref{app:CoTS-sdft-u}.
To construct the Wannier functions, the inner window to fix the low energy band dispersion was set from -8 to  1.5 eV, and a 6$\times$6$\times$3 sampling $k$-point grid was used.
Here, the 3$d$ orbitals of Co, 5$d$ orbitals of Ta, and $p$ orbitals of S atoms were employed to fit the \textit{ab initio} band structure.

For \PES, we set the lattice constants as the experimental value of $a = 9.2524$ \AA~\cite{Madsen2001-rr}.
The energy cut-off for the plane-wave basis set was set to 600 eV, and we used a 4$\times$4$\times$4 $k$-point grid for the primitive cell.
In the pseudopotentials, the valence electron configurations were 3$p^6$3$s^1$ (K), 3$s^2$3$p^1$ (Al), 3$s^2$3$p^2$ (Si), and 2$s^2$2$p^4$ (O), respectively.
Following Nakamura \textit{et al}~\cite{Nakamura2009-th}, we did not apply SDFT$+U$ calculation for this compound.
To construct the Wannier functions, the inner window to fix the low energy band dispersion was set from -1 to  1 eV, and a 4$\times$4$\times$4 sampling $k$-point grid was used.
For this compound, two $s$ orbitals at (0, 0, 0) and (0.5, 0.5, 0.5) in units of the primitive lattice vectors were employed to fit the \textit{ab initio} band structure.
It should be noted that these $s$ orbitals reside in the interstitial region, without being localized at any specific atomic site.
This enabled the downfolding of the electronic structure into a simple model consisting of only two orbitals per unit cell.

%------------------------------------------------------------------------
%------------------------------------------------------------------------
%------------------------------------------------------------------------
%------------------------------------------------------------------------
\section{Benchmarks for the Minimal Model and Elemental Magnetic Metals}\label{sec:benchmark}
%------------------------------------------------------------------------
%------------------------------------------------------------------------

%------------------------------------------------------------------------
%------------------------------------------------------------------------
\subsection{Two-sublattice Hubbard Model}\label{subsec:2-orb}

\begin{figure}[htbp]
    \centering
    \includegraphics[keepaspectratio, width=0.7\columnwidth]{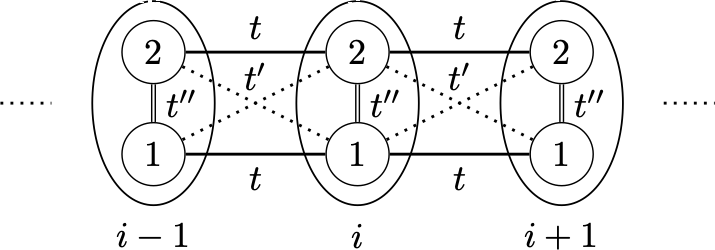}
    \caption{One-dimensional single-orbital Hubbard model with two sublattices. The Hamiltonian has three types of transfer integrals, $t$, $t'$, and $t''$, defined for the pairs of sublattices shown with the solid, dashed, and double lines, respectively.}\label{fig:tnk-model}
\end{figure}

Here, we first study a one-dimensional single-orbital Hubbard model with two sublattices considered in Ref.~\cite{tanaka}. This model offers the simplest case that exhibits the BQ interaction when deriving an effective quantum spin model.
The Hamiltonian is defined as
\begin{align}
    \mathcal{H} &= -\sum_{\langle i\ell',jm'\rangle,\sigma}(t_{i\ell',jm'}\hat{c}^{\dagger}_{i\ell'\sigma}\hat{c}_{jm'\sigma}+\mathrm{h.c.})+\sum_{i}U\hat{n}_{i\uparrow}\hat{n}_{i\downarrow}\label{eq:hubbard-ham}
\end{align}
where $(i, j), (\ell', m)$ and $\sigma$ are the degrees of freedom of lattice, sublattice, and spin, respectively.
The bracket $\langle\rangle$ stands for the summation of the combinations between neighboring lattices.
$\hat{c}_{i\ell'\sigma}/\hat{c}^{\dagger}_{i\ell'\sigma}$ and $ \hat{n}_{i\sigma}$ are the annihilation/creation and number operator of an electron.
We show the schematic picture of the model in Fig.~\ref{fig:tnk-model}.

The Hubbard Hamiltonian is defined using the spin-1/2 degree of freedom for each sublattice.
When two sublattices form a dimer and, consequently, a triplet, we can derive an effective quantum spin model with the spin-1 degree of freedom for each lattice in the limit of strong correlation as follows~\cite{tanaka}%}
:
\begin{align}
    \mathcal{H}_{\mathrm{eff}} &= -2\qty[\sum_{\langle i,j\rangle}J^{\mathrm{Q}}_{ij}(\hat{\bm{S}}_i\cdot\hat{\bm{S}}_j)+\sum_{\langle i,j\rangle}B^{\mathrm{Q}}_{ij}(\hat{\bm{S}}_i\cdot\hat{\bm{S}}_j)^2],
\end{align}
where the superscript Q denotes the interactions in the quantum spin model. The BL and BQ interactions can be obtained perturbatively as follows~\cite{tanaka}:
\begin{align}
    J^{\mathrm{Q}}_{ij} = -\frac{t^2+t'^2}{U},\ \ B^{\mathrm{Q}}_{ij} = -\frac{20t^2t'^2}{U^3}.\label{eq:jij-pert}
\end{align}

Next, to compare the interactions in the quantum spin model with those in the classical spin model evaluated from SCE-DLM, denoted as $J^{\mathrm{C}}_{ij}$ and $B^{\mathrm{C}}_{ij}$, respectively, we take the classical limit of the quantum spin~\cite{Lieb2004}.
For an $S$-spin length, the interactions need to be rescaled as follows~\cite{rescale}:
\begin{align}
    J^{\mathrm{Q}}_{ij}(\hat{\bm{S}}_i\cdot\hat{\bm{S}}_j)&\rightarrow S^{2}J^{\mathrm{C}}_{ij}(\bm{e}_i\cdot\bm{e}_j)\label{eq:rescale-jij}\\
    B^{\mathrm{Q}}_{ij}(\hat{\bm{S}}_i\cdot\hat{\bm{S}}_j)^2&\rightarrow S^{4}B^{\mathrm{C}}_{ij}(\bm{e}_i\cdot\bm{e}_j)^2.\label{eq:rescale-bij}
\end{align}
Hence, we compare $J^{\mathrm{C}}_{ij}$ with $J^{\mathrm{Q}}_{ij}$, and likewise, $B^{\mathrm{C}}_{ij}$ with $B^{\mathrm{Q}}_{ij}$, since we are examining the spin-1 case, where $S=1$.
SCE-DLM calculates spin interactions between a lattice consisting of two sublattices.

To apply SCE-DLM to this model, we first construct a tight-binding Hamiltonian including both the hopping parameter $t$ and spin splitting $B$, which can be obtained via the mean-field approximation for the half-filled state of the Hamiltonian Eq.~(\ref{eq:hubbard-ham}):
\begin{align}\label{eq:mf-ham}
    \mathcal{H}_{\mathrm{MF}} =& -\sum_{\langle i\ell',jm'\rangle,\sigma}(t_{i\ell',jm'}\hat{c}^{\dagger}_{i\ell'\sigma}\hat{c}_{jm'\sigma}+\mathrm{h.c.}) - \bm{B}_i\cdot\hat{\bm{m}}_i\\
    \bm{B}_i &= \frac{U}{2}\expect{\hat{\bm{m}}_i},\quad \hat{\bm{m}}_i = \sum_{\ell',\sigma,\sigma'}\hat{c}^{\dagger}_{i\ell'\sigma}\bm{\sigma}\hat{c}_{i\ell'\sigma'}\label{eq:mf-split}
\end{align}
For the half-filled case with $t\ll U$, the magnetization operator $\hat{\bm{m}}_i$ becomes $\sigma_z\hat{\bm{z}}$.
Consequently, we can obtain the Hubbard parameter $U$ from the magnitude of the spin splitting $B=|\bm{B}|$, i.e., $U = 2B$.

\begin{figure}[htbp]
    \centering
    \includegraphics[width=\columnwidth, keepaspectratio]{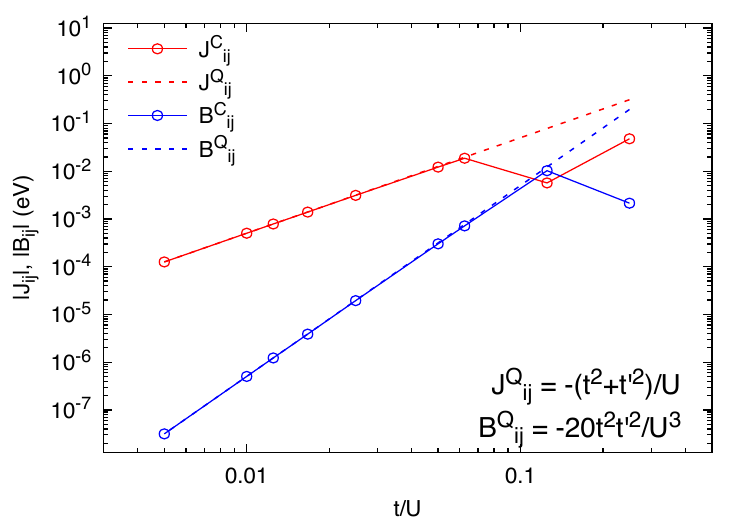}
    \caption{BL ($J^{\mathrm{C}}_{ij}/J^{\mathrm{Q}}_{ij}$) and BQ ($B^{\mathrm{C}}_{ij}/B^{\mathrm{Q}}_{ij}$) interactions in the classical/quantum spin Hamiltonian for $t=t'=t''$ in the limit of strong correlation $t\ll U$. We set $B=U/2=10$ in the calculations. $J^{\mathrm{C}}_{ij}$ (solid red line) and $B^{\mathrm{C}}_{ij}$ (solid blue line) are evaluated by SCE-DLM, and $J^{\mathrm{Q}}_{ij}$ (dashed red line) and $B^{\mathrm{Q}}_{ij}$ (dashed blue line) are evaluated perturbatively (see Eq.~(\ref{eq:jij-pert})). }
    \label{fig:tnk-conv-tu}
\end{figure}

In Fig. \ref{fig:tnk-conv-tu}, we plot $J^{\mathrm{C}}_{ij}, J^{\mathrm{Q}}_{ij}, B^{\mathrm{C}}_{ij}, B^{\mathrm{Q}}_{ij}$ for the half-filled case as a function of $t/U$.
{Following Ref.~\cite{tanaka}, we take a $t=t'=t''$ case.}
% It is worth noting that the chemical potential $\mu_c$ of the DLM state is always zero for the half-filled state.
We can see that the interactions evaluated perturbatively for the quantum spin model align closely with those derived from SCE-DLM in the limit of $t/U\rightarrow 0$.
This result suggests that the present method is applicable to a wide variety of strongly correlated magnetic compounds with a strong local moment.

In appendix \ref{app:asymp-sorb}, we give the asymptotic form of spin interactions derived from our approach applied to a single-orbital and single-sublattice Hubbard model.
It is shown that, regarding the BL interaction, the asymptotic forms are the same as those of the LKAG method in both the strongly correlated and itinerant limits.
Additionally, in Appendix \ref{app:comp-lkag}, we present a comparison with the LKAG method concerning the dependence on the initial magnetic order.
In this comparison, we applied our method to a single-orbital model on the square lattice.
The results demonstrate that the SCE-DLM is certainly less dependent on the initial magnetic order than the LKAG method.

%------------------------------------------------------------------------
%------------------------------------------------------------------------

\subsection{BCC Fe}\label{subsec:bcc-fe}
\begin{figure*}[htbp]
    \centering
    \includegraphics[width=1.8\columnwidth, keepaspectratio, page=1]{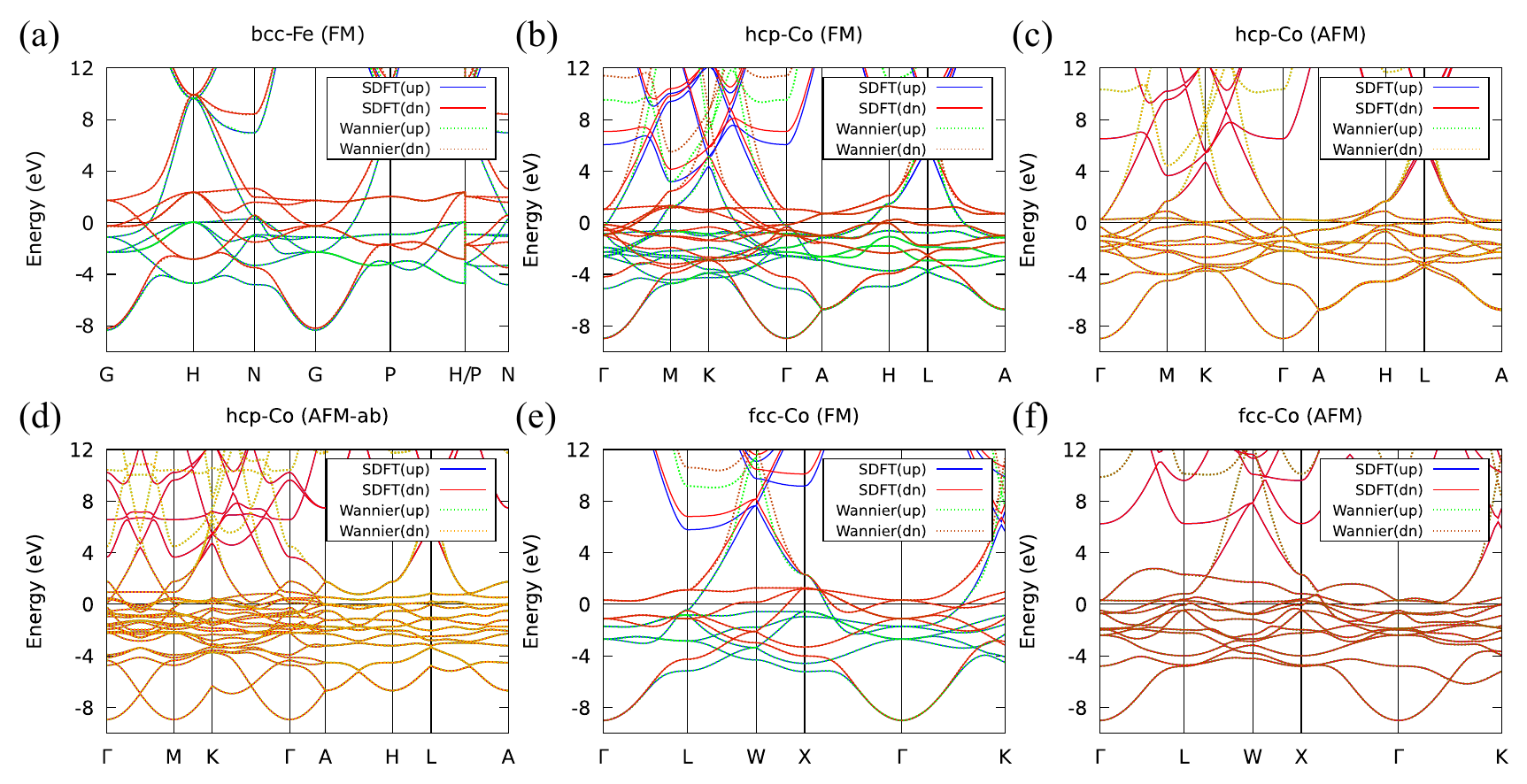}
    \caption{Band structure of (a) the FM state of bcc Fe. (b)-(d) are those of the FM, AFM, and AFM-$ab$ states of hcp Co. (e) and (f) are those of the FM and AFM states of fcc Co. Blue and red lines are those for the up and down spin components obtained by SDFT calculation, and green and orange lines are those obtained by Wannier interpolation. The labels of the Brillouin zone are defined based on the primitive cell of each lattice.}\label{fig:bands}
\end{figure*}

Next, we apply our scheme to the \textit{ab initio} tight-binding models for a prototypical magnetic metal, bcc Fe.
In Fig.~\ref{fig:bands} (a), we present the band structures of these magnets obtained by SDFT calculations and those fitted by the Wannier-based tight-binding model for bcc Fe.

\begin{figure}[htbp]
    \centering
    \includegraphics[width=0.8\columnwidth, keepaspectratio, page=1]{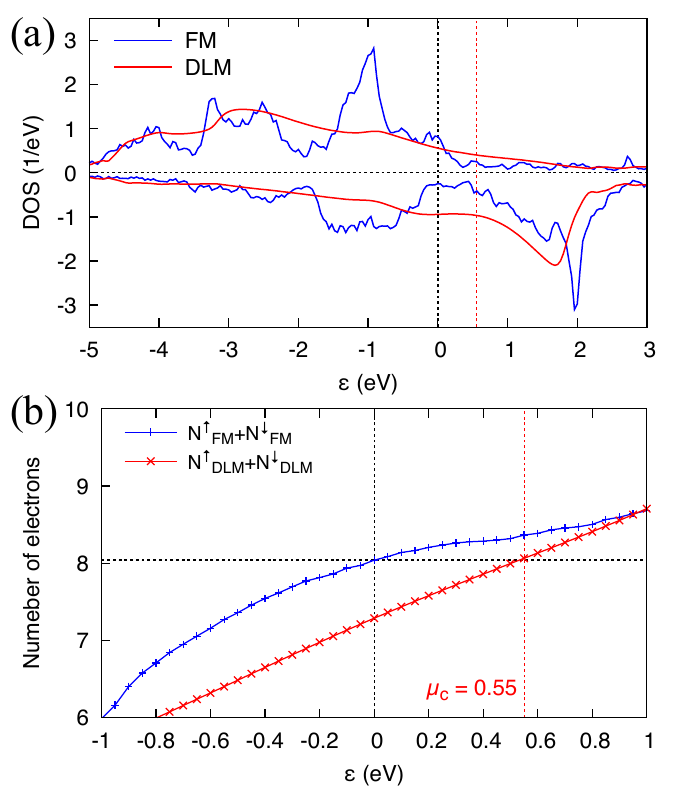}
    \caption{Density of states (DOS) for the spin-up and spin-down components (a) and the integrated DOS (b) for bcc Fe.
    In panel (a), the blue (red) line represents the DOS for the FM (DLM) state.
    The spin-up and spin-down components of both the FM and DLM states are plotted on the positive and negative sides, respectively.
    In panel (b), the vertical black line indicates the chemical potential of the FM state ($\mu=0$), and the vertical red line indicates the chemical potential $\mu_c$ for the DLM state.
    The horizontal line indicates the number of electrons at $\mu=0$ in the \textit{ab initio} tight-binding Hamiltonian.}\label{fig:dlm-dos}
\end{figure}

In Fig. \ref{fig:dlm-dos}, we show the density of states (DOS) and integrated DOS  along with the calibrated chemical potential $\mu_c$ for the DLM state.
Let us now compare the chemical potential and magnetic moment of the DLM and ferromagnetic (FM) state.
Following the procedure outlined in Refs.~\cite{Gyorffy1972-nb, wannier-cpa}, we calculated the DOS for each spin component of the DLM state using the Green's function:
\begin{align}
    \gbar^{\sigma\sigma}_{ii} = \gbar_{ii}+\gbar_{ii}T_{i}(\sigma\hat{\bm{z}})\gbar_{ii}. \label{eq:dlm-project}
\end{align}
$\mu_c$, the chemical potential of the DLM state measured from that of the FM state, is 0.55 eV.
Namely, $\mu$ of bcc Fe strongly depends on the changes in the electronic/magnetic structure.
Regarding the magnetic moment, which is defined as the difference in the number of spin-up and spin-down electrons up to the chemical potential, it is 2.27 (2.28) $\mu_{\textrm{B}}$ for the FM (DLM) state.

\begin{figure}
    \centering
    \includegraphics[keepaspectratio, width=0.9\columnwidth]{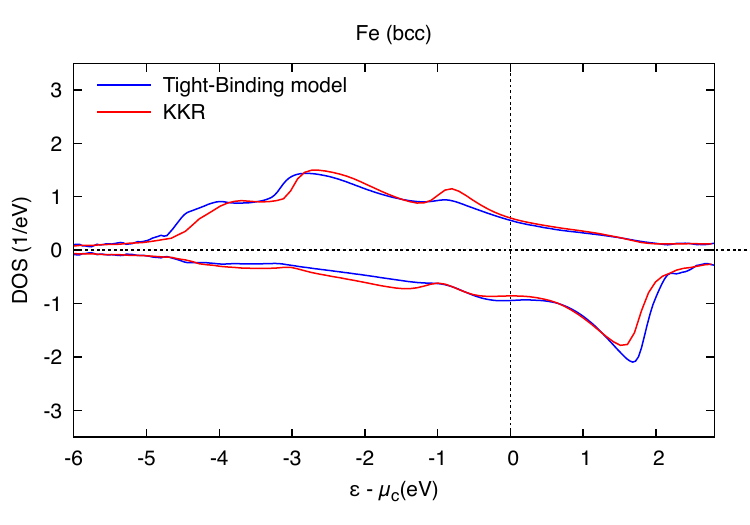}
    \caption{Density of states of the DLM states obtained by the \textit{ab initio} tight-binding method (blue) and the KKR method (red). The energy axis is calibrated by subtracting $\mu_c$ from each result.}
    \label{fig:dos-comp}
\end{figure}

To verify the validity of these results, we compare the DLM state for bcc Fe obtained by the \textit{ab initio} tight-binding (TB) method and that calculated by the KKR method.
In the KKR calculation, the magnetic moment is calculated to be 2.21 (2.18) $\mu_{\textrm{B}}$ for the FM (DLM) state, and the Fermi energy of the DLM state relative to the FM state is 0.33 eV.
Similar to the result for the tight-binding model, while the magnitude of the magnetic moment hardly depends on the magnetic states, the Fermi energies for the FM and DLM states are quite different.
In Fig. \ref{fig:dos-comp}, we compare the density of states of the DLM state obtained using the KKR method with that in Fig. \ref{fig:dlm-dos} (a).
Note that the energy axis in Fig. \ref{fig:dos-comp} is adjusted by subtracting $\mu_c$ for each method.
We see that the density of states obtained from these two methods are closely aligned.

\begin{figure}[htbp]
    % \begin{minipage}[b]{0.9\columnwidth}
    \centering
    \includegraphics[width=\columnwidth, keepaspectratio, page=1]{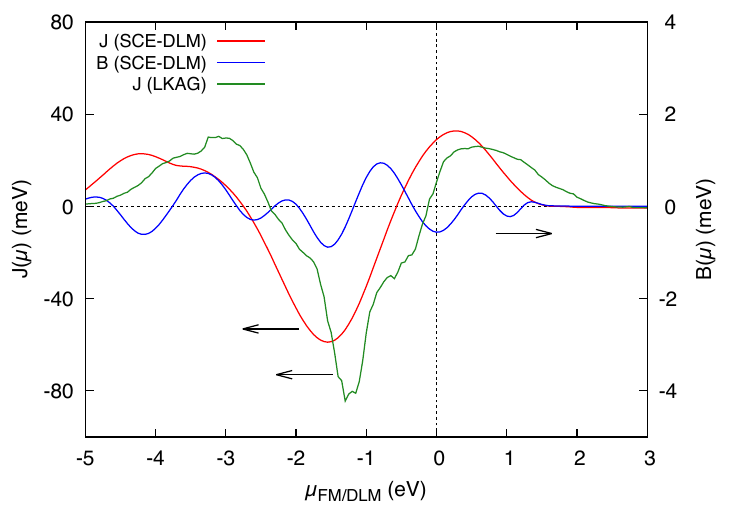}
    \caption{Chemical potential dependence of the nearest-neighbor BL ($J$) and BQ ($B$) interactions for bcc Fe.
    Red, blue, and green lines indicate $J$ and $B$ calculated by SCE-DLM and $J$ calculated by LKAG, respectively. Refer to the left vertical axis for $J$, and to the right vertical axis for $B$.
    }\label{fig:jij-edep-tm3d}
\end{figure}

We then evaluate the nearest-neighbor BL ($J$) and BQ ($B$) interactions by SCE-DLM.
In Fig.~\ref{fig:jij-edep-tm3d}, we plot $J$ and $B$ as a function of the chemical potential $\mu$.
For $J$, we compare the result with that obtained by the LKAG approach.
We here define $\mu_{\textrm{FM}} = \mu$ for the LKAG approach and $\mu_{\textrm{DLM}} = \mu-\mu_c$ for the SCE-DLM approach.
It can be seen that the dependence on $\muFM$ in LKAG and the dependence on $\muDLM$ in SCE-DLM are similar.
It should be noted that this $\muFM$ dependence in the LKAG qualitatively explains the magnetism observed in 3$d$ transition metals~\cite{sakuma-3d}.
Regarding $B$, we see that its energy scale is much smaller than that of $J$ since it is a higher-order interaction involving more scattering processes. Though, $B$ shows many sign changes as a function of $\muDLM$, it is negative at $\muDLM=0$.
 
\begin{figure*}[htbp]
    \centering
    \includegraphics[width=1.8\columnwidth, keepaspectratio]{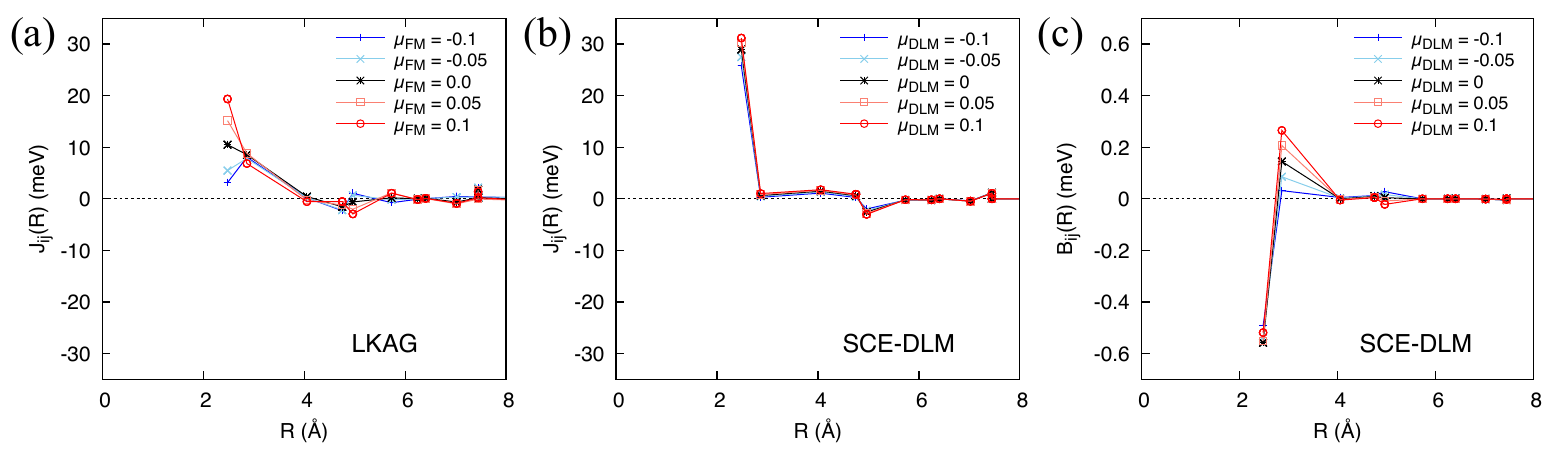}
    \caption{The distance dependence of the BL ($J_{ij}$) and the BQ ($B_{ij}$) interactions. $J_{ij}$ calculated by LKAG is shown in the panel (a). $J_{ij}(B_{ij})$ calculated by SCE-DLM are shown in the panel (b)((c)). We also plot the values when we change $\mu_{\textrm{FM/DLM}}$.}\label{fig:jij-rdep-tm3d}
\end{figure*}

\begin{figure}
    \centering
    \includegraphics[width=\columnwidth, keepaspectratio, page=1]{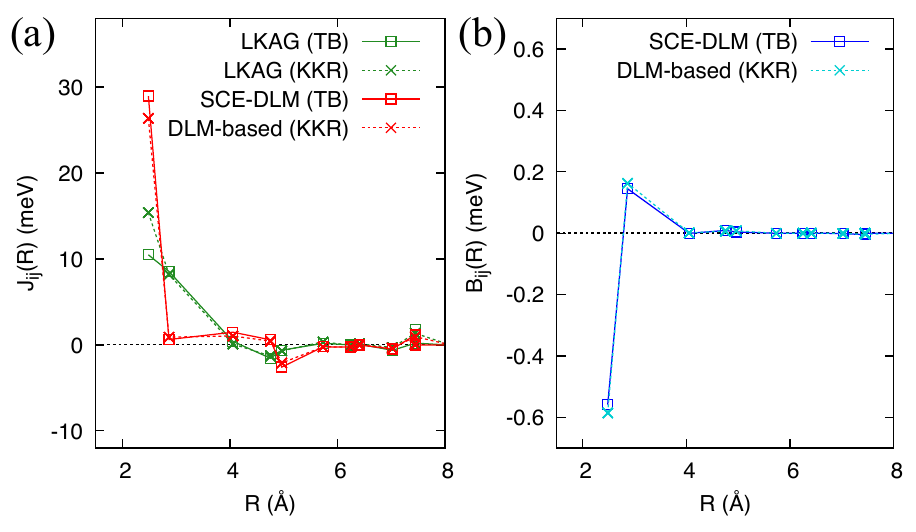}
    \caption{The (a) BL and (b) BQ interactions of bcc Fe as a function of the distance ($R$) between the $i$-th and $j$-th site. The green (red and blue) lines are those calculated using LKAG (SCE-DLM) at $\mu_{\textrm{FM(DLM)}}=0$. Among the same color lines, solid (dotted) lines are the results of the \textit{ab initio} tight-binding (KKR) method.
    }
    \label{fig:fe-jij-rdep-comp-kkr}
\end{figure}

In Fig.~\ref{fig:jij-rdep-tm3d}, we plot the BL ($J_{ij}$) and BQ ($B_{ij}$) interactions as a function of the distance ($R$) between the $i$-th and $j$-th site for the range of $-0.1\leq\mu_{\textrm{FM/DLM}}\leq 0.1$ for both LKAG and SCE-DLM methods.
From Figs. ~\ref{fig:jij-rdep-tm3d}(a) and (b), we see that the $\mu_{\textrm{FM}}$ dependence around $\mu_{\textrm{FM}}=0$ of the NN interaction $J$ is significant for bcc Fe in the LKAG calculation, which could cause sizable computational errors in the evaluation of $J$.
In addition, we observe the following from Fig.~\ref{fig:jij-rdep-tm3d}:
(a, b) For bcc Fe, the second NN interaction calculated by LKAG is as large as the NN interaction, and the one calculated by SCE-DLM is negligibly small.
These features align with the results using the KKR method as shown in Fig. \ref{fig:fe-jij-rdep-comp-kkr} (a).
In Fig.~\ref{fig:jij-rdep-tm3d}~(c), for bcc Fe, the size of the second NN $B_{ij}$ is as large as that of the NN interaction, which also aligns with the results of the approach based on fitting the spin-configuration dependence of the total energy~\cite{Jacobsson2022-nk}.
The $\muDLM$ dependence of the NN interaction $\muDLM$ of $B_{ij}$ for the NN interaction is weak, while the second NN neighbour changes considerably.

In Figs.~\ref{fig:fe-jij-rdep-comp-kkr} (a) and (b), we show the results of $J_{ij}$ and $B_{ij}$ as a function of the distance $R$ between sites $i$ and $j$ at $\mu_{\textrm{FM/DLM}}=0$ for bcc Fe.
We compare the magnetic interactions evaluated by the four methods: (1) LKAG for the tight-binding model, (2) LKAG within the KKR scheme, (3) SCE-DLM for the tight-binding model,  and (4) the local force approach for the DLM state within the KKR scheme.
On the one hand, for the LKAG methods (method $(1)$ and $(2)$), the computed $J_{ij}$'s differ by about $5$ meV, which can be attributed to the dependence on the implementation details~\cite{Szilva2023-lr}, and the strong dependence on the chemical potential (as shown Fig.~\ref{fig:jij-rdep-tm3d} (a)).
On the other hand, $J_{ij}$ evaluated employing the DLM-based methods (method (3) and method (4)) closely agree with each other up to the long-range limit.
In addition, the BQ interactions $(B_{ij})$ computed using the two DLM-based approaches  (method (3) and method (4)) are very close.
For bcc Fe in the DLM state, the dependence on the chemical potential is weak, resulting in a better agreement between the tight-binding and KKR methods.

Our approach, based on the tight-binding model, differs from the KKR method in terms of the dependence on the starting magnetic order, as mentioned in Sec. \ref{subsec:dlm}.
However, it is noteworthy that these two distinct methods yield nearly identical DLM states and spin interactions for magnets where local magnetic moments are well-defined, such as bcc Fe.

A variety of methods to calculate the BQ interaction from first principles have been proposed. So far, theoretical calculations for bcc Fe have yielded both positive\cite{extension, sce_Fe} and negative values\cite{Freeman1995-gj, Spisak1997-jo, szilva, Jacobsson2022-nk, tb2j} for the nearest-neighbor BQ interaction.
Aside from Refs.~\cite{Jacobsson2022-nk, sce_Fe}, these works are based on LKAG and rely on the FM reference state to evaluate the BQ interaction.
Our SCE-DLM approach differs from such approaches in that it calculates spin interactions from a disordered state, specifically the DLM state, rather than an ordered state.
As is outlined in Ref. \cite{Dos_Santos_Dias2022-ec}, approaches with a magnetically ordered state are valid for calculating physical properties related to the specific ordered state, such as the magnon spectrum~\cite{szilva}.
In contrast, approaches without an ordered reference state are better suited for exploring arbitrary magnetic configurations, e.g., when constructing a phase diagram of the system.

The approaches based on the fitting of the spin-configuration dependence of the total energy can also yield different results~\cite{Jacobsson2022-nk, sce_Fe}.  
In Ref. \cite{Jacobsson2022-nk}, they fit the spin Hamiltonian to the \textit{ab initio} energies for a number of spin spiral states with random wave vectors.
Though they confine the spin Hamiltonian up to the BQ interaction and a four-spin interaction, they do not prioritize interactions between specific pairs during the fitting.
On the other hand, in Ref. \cite{sce_Fe}, while considering arbitrary spin interactions in the spin cluster expansion, they estimate the nearest-neighbor BQ interaction after the nearest-neighbor BL interaction. 
These different approaches could lead to a difference in the sign of the calculated BQ interactions. 
Additionally, as illustrated in Figs. \ref{fig:jij-rdep-tm3d} (a)-(f), the chemical potential dependence of the spin interactions could introduce ambiguity in theoretical results.
Our approach is similar to the approach employed in Ref.~\cite{Jacobsson2022-nk} regarding the accessibility to various spin configurations, and our results are consistent with their result.

\begin{figure*}
    \centering
    \begin{minipage}[b]{0.9\columnwidth}
        \centering
        \includegraphics[width=\columnwidth, keepaspectratio, page=1]{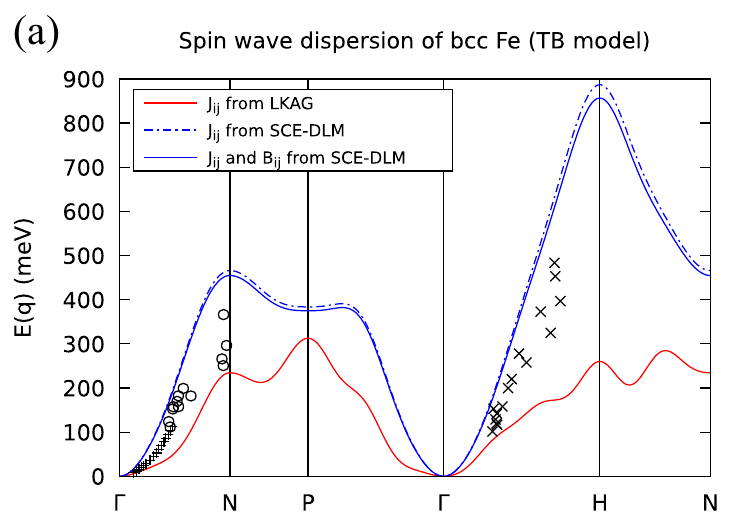}
    \end{minipage}
    \hspace{0.04\columnwidth}
    \begin{minipage}[b]{0.9\columnwidth}
        \centering
        \includegraphics[width=\columnwidth, keepaspectratio, page=2]{Fig9_fe_bcc_magnon_band.pdf}
    \end{minipage}
    \caption{
    Spin-wave dispersion of bcc Fe. The interactions are obtained (a) based on the tight-binding model and (b) using the KKR method. The red curve is the dispersion calculated from the BL interactions using LKAG. The blue dashed (solid) curves are those from the BL interactions only (and the BQ interactions) using SCE-DLM. The dots in the figure are experimental results ($+$: Ref.\cite{Lynn1975-td}, $\times$: Ref.\cite{Perring1991-px}, $\textcircled{}$: Ref.\cite{Boothroyd1992-ke}).
    }
    \label{fig:spin-wave}
\end{figure*}

As noted previously, the appropriate physical quantities differ depending on whether the calculations are conducted with or without a magnetically ordered state.
To highlight this distinction, we analyze the spin-wave dispersion in the Brillouin zone of the BCC lattice, comparing results obtained using LKAG and SCE-DLM.
In Figs. \ref{fig:spin-wave} (a) and (b), we show the theoretical spin-wave dispersion calculated from $J_{ij}$ using LKAG, $J_{ij}$ using SCE-DLM, $J_{ij}$ and $B_{ij}$ using SCE-DLM combined, along with the experimental results~\cite{Lynn1975-td, Perring1991-px, Boothroyd1992-ke}.
The dispersion $E(\bm{q})$ is computed via~\cite{Iwashita1976-ih, Pajda2001-yt}:
\begin{align}\label{eq:spin-wave}
    E(\bm{q}) = \frac{4\mu_B}{M}\sum_{j\neq{i}}\qty(J_{ij}+2B_{ij})(1-e^{i\bm{q}\cdot\bm{r}_{ij}}),
\end{align}
where $\mu_B$ and $M$ refer to the Bohr magneton and the magnitude of the magnetic moment, respectively.
In LKAG, the spin interactions are derived from the second-order derivative of the energy. Hence, the calculated $J_{ij}$ inherently includes contributions from higher-order interactions\cite{Spisak1997-jo}, which indicates that $B_{ij}=0$ for the linear spin-wave calculation using LKAG.
In contrast, the SCE-DLM method treats the BL and BQ interactions separately with distinct basis functions.
Thus, the contribution from the BQ interaction must be added explicitly to the BL interaction.

In Fig. \ref{fig:spin-wave} (a), it is observed that the spin-wave dispersion calculated using LKAG is underestimated compared to the experimental results.
However, LKAG shows good agreement with experimental values near the $\Gamma$ point when using the KKR (see Fig. \ref{fig:spin-wave} (b)) or the tight-binding linear muffin-tin orbital (TB-LMTO) method~\cite{Pajda2001-yt}.
One reason for the discrepancy in our calculation is that, as shown in Fig. \ref{fig:fe-jij-rdep-comp-kkr}, the first NN $J_{ij}$ obtained from our tight-binding model is smaller than those obtained from KKR.
$J_{ij}$ calculated using KKR or TB-LMTO also exhibits some variation depending on implementations~\cite{Szilva2023-lr}.
Additionally, as shown in Fig. \ref{fig:jij-rdep-tm3d} (a), the first NN $J_{ij}$ evaluated by LKAG is highly sensitive to the chemical potential~\cite{sakuma-3d, local-force-nmt}.
Thus, this difference in the NN interactions contributes to the underestimation of the spin-wave dispersion in the LKAG calculation for our tight-binding model.

In contrast, the spin-wave dispersion calculated using only the BL interactions from SCE-DLM tends to be overestimated compared to the experimental results, especially along the $\Gamma-$H line.
Upon including the BQ interaction, the spin-wave dispersion is lowered slightly but non-negligibly; in our result, the NN ratio $2B/J\sim{0.04}$, indicating a few-percent downward shift.
This shows that the BQ term also affects the spin-wave dispersion in ferromagnets.

As for the residual discrepancy, several factors are plausible. (i) Omitted interactions. Within SCE–DLM, distinct higher-order interactions contribute separately to the magnon dispersion, whereas LKAG effectively folds these effects into the BL terms; the remaining mismatch can thus be ascribed to contributions neglected in our calculation. (ii) Magnetic anisotropy. In bcc Fe, its energy scale is small, and its impact away from the $\Gamma$ point is limited, so its overall effect is expected to be weak. Though a systematic inclusion of these terms to fully converge the SCE–DLM dispersion is, in principle, possible, it would require substantial computational effort and is beyond the scope of the present study.

A further source of discrepancy stems from the single-site CPA underpinning the DLM framework: long-range correlations in the DLM state are entirely neglected, which can influence the extracted spin interactions and hence the dispersion. In addition, the classical-spin approximation (also used in LKAG) may contribute to quantitative differences.
While experimental data for the spin-wave dispersion in high-energy regions away from the $\Gamma$ point are still limited, it is expected that these calculations will be quantitatively verified in future studies.

%------------------------------------------------------------------------
%------------------------------------------------------------------------
\subsection{HCP Co}\label{subsec:hcp-co}

Although our approach does not entirely eliminate the dependence on the starting magnetic order, we find that using the tight-binding model of the FM state results in spin interactions nearly identical to those obtained by the KKR method for magnets with well-defined local magnetic moments, such as bcc Fe.
However, this agreement is not always maintained for magnets with more itinerant magnetic moments.

\begin{figure*}[htbp]
    \centering
    \includegraphics[width=1.8\columnwidth, keepaspectratio, page=1]{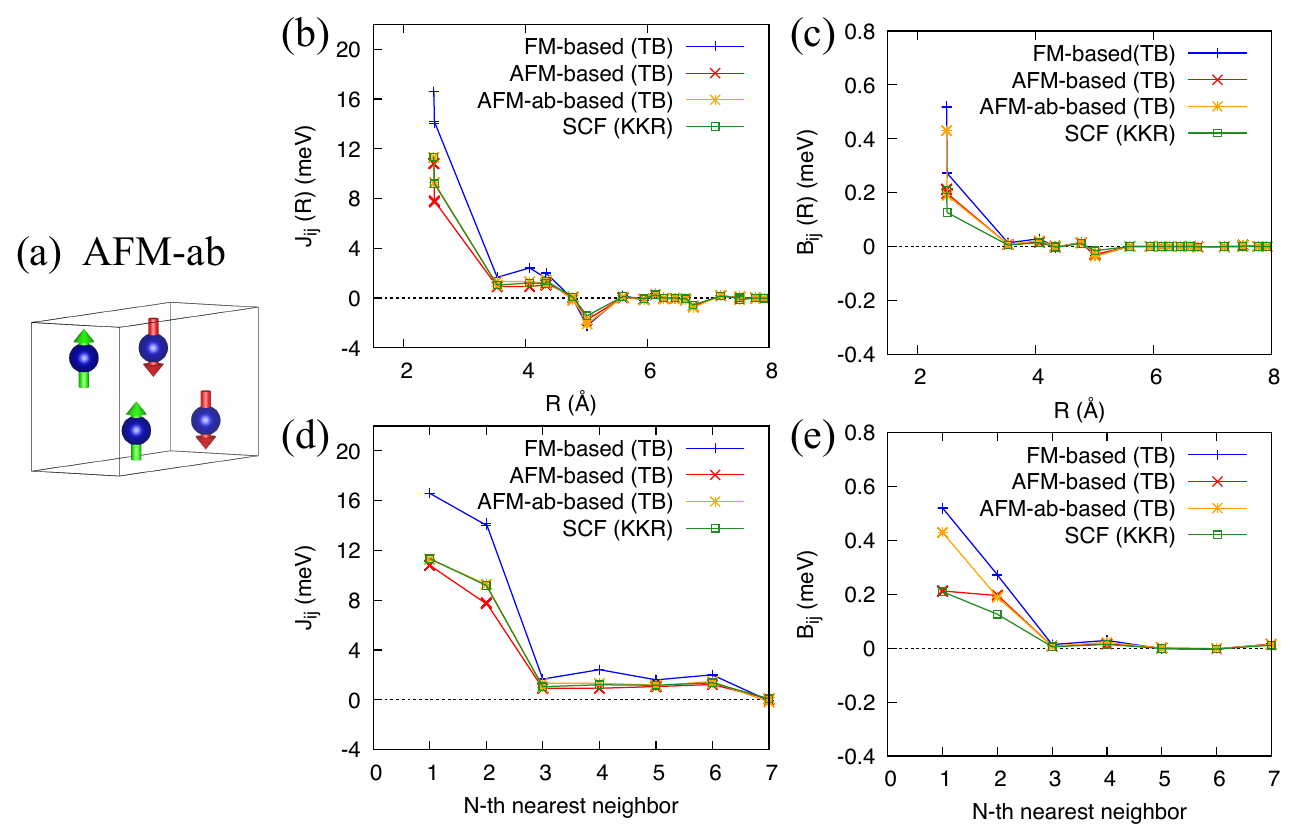}
    \caption{(a) Magnetic structure of the AFM-$ab$ state. (b) BL and (c) BQ interactions of hcp Co as a function of the distance ($R$) between the $i$-th and $j$-th site. The blue, red, orange, and green lines are those calculated using the DLM states based on the FM, AFM, and AFM-$ab$ states in our approach, and the self-consistent KKR method, respectively. In panels (d) and (e), we plot the same interactions in the order of $N$-th NNs for better visualization.}\label{fig:coh-jij-rdep-comp-kkr}
\end{figure*}

We compare the results of our approach, starting from the FM state, with those obtained from the KKR method for hcp Co.
Additionally, we consider two antiferromagnetic (AFM) states: one corresponds to a configuration in which two magnetic moments within the primitive cell of the HCP lattice are oriented in opposite directions, while the other is the magnetic structure shown in Fig.~\ref{fig:coh-jij-rdep-comp-kkr} (a), referred to as AFM-$ab$.
In Fig.~\ref{fig:bands} (b)-(d), we present the band structures of these states, obtained through SDFT calculations and fitted using the Wannier-based tight-binding model for hcp Co.
Hereafter, we refer to the results as FM-based, AFM-based, and AFM-$ab$-based, whereas those obtained using the KKR method are denoted as self-consistent field (SCF)-DLM.
Table~\ref{tb:coh-comp-momsize} summarizes the magnitudes of the magnetic moments calculated by these methods.

\begin{table}[htbp]
    \renewcommand{\arraystretch}{1.7}
    \centering
    \begin{tabular}{c|c|c|c|c}%{C{2.6cm}|C{1cm}|C{1.2cm}|C{1.3cm}|C{1.8cm}}
         & KKR & FM-based & AFM-based & AFM-$ab$-based\\ \hline\hline
        Ordered state ($\muB$) & 1.62 & 1.61 & 1.06 & 1.20\\ \hline
        DLM state ($\muB$) & 1.19 & 1.44 & 1.10 & 1.24
    \end{tabular}
    \caption{Magnitudes of the magnetic moments ($\muB$) of the ordered states and the DLM states obtained from the FM, AFM, and AFM-$ab$ states, and those calculated using the KKR method.}
    \label{tb:coh-comp-momsize}
\end{table}

In Fig. \ref{fig:coh-jij-rdep-comp-kkr}, we show the spin interactions calculated from the four DLM states as a function of distance.
To improve visualization, the interactions are also presented in terms of the $N$-th NN order, as the distances to the first and second NNs are very close.
For the AFM-$ab$ result, since the interactions vary between equivalent Co-Co bonds, we plot their averaged values.
These discrepancies arise from the spin-dependent hopping terms in the TB model due to the symmetry breaking in the reference magnetic order.
The effects of these terms were on the order of $O(10^{-1} \textrm{meV})$ for hcp Co and are subtle for the BL interactions.
The magnitude of these terms is material-dependent and should be examined when starting from a complex magnetic structure~\cite{local-force-nmt}.
Since these terms stem from inter-site hopping, they diminish as the relevant spin-dependent contributions become more localized (for instance, with more localized magnetic moments).
As demonstrated for Co, its impact is expected to be practically small within the applicability range of our approach.

For hcp Co, the FM-based and KKR results differ noticeably from each other, in contrast to the case of bcc Fe.
This discrepancy arises because the FM-based DLM state has larger magnetic moments than the SCF-DLM state obtained by the KKR method.
In contrast, the AFM- and AFM-$ab$-based results yield magnetic moments much closer to the KKR values and show better agreement in spin interactions.
This suggests that the dependence on the starting magnetic order is weaker among states with closer magnitudes of magnetic moments.
Additionally, when the magnitude of the magnetic moments is similar to that in the SCF-DLM state, spin interactions calculated by our approach closely align with those obtained from the SCF-DLM state.

\begin{figure}[htbp]
    \centering
    \includegraphics[width=\columnwidth, keepaspectratio, page=1]{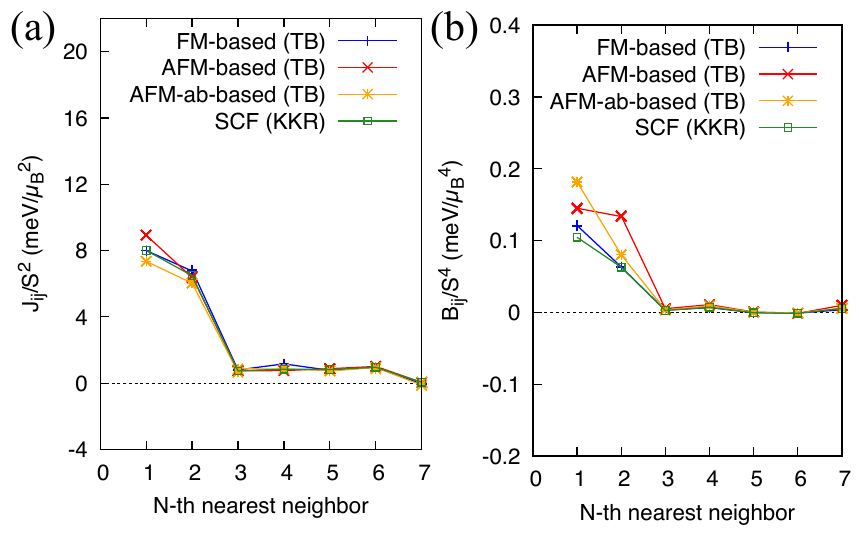}
    \caption{(a) BL and (b) BQ interactions rescaled with the magnitude of magnetic moments of hcp Co in the order of $N$-th NNs. The blue, red, orange, and green lines are those calculated using the DLM states based on the FM, AFM, and AFM-$ab$ states in our approach, and the KKR method, respectively.}\label{fig:coh-jij-rdep-comp-kkr-rescale}
\end{figure}

In our approach, the magnitude of magnetic moments is incorporated into spin interactions, since our approach defines spin interactions within a spin Hamiltonian where classical spins are described by unit vectors.
As a result, BL and BQ interactions are rescaled by $S^2$ and $S^4$, respectively, with $S$ being the magnitude of the magnetic moments.
To explore the effect of the magnitude of magnetic moments, we present spin interactions rescaled with the magnitudes of magnetic moments, i.e. $J_{ij}/S^2$ and $B_{ij}/S^4$, in Fig.~\ref{fig:coh-jij-rdep-comp-kkr-rescale}.
The rescaled BL interactions show good agreement across different initial magnetic orders.
In the case of the BQ interaction, though the AFM- and AFM-$ab$-based results are not in perfect agreement, the rescaled FM-based result is closer to the KKR value than the raw interactions in Fig.~\ref{fig:coh-jij-rdep-comp-kkr} (c) and (e).

From these results, we can see that spin interactions calculated by our approach show weak dependence on the starting magnetic order when the magnitudes of magnetic moments are close.
Even when the magnitudes are not exactly the same, the magnitude of magnetic moments allows us to estimate the range of the values of spin interactions.

%------------------------------------------------------------------------
%------------------------------------------------------------------------
\subsection{FCC Co}\label{subsec:fcc-Co}

\begin{figure}[htbp]
    \centering
    \includegraphics[width=\columnwidth, keepaspectratio, page=1]{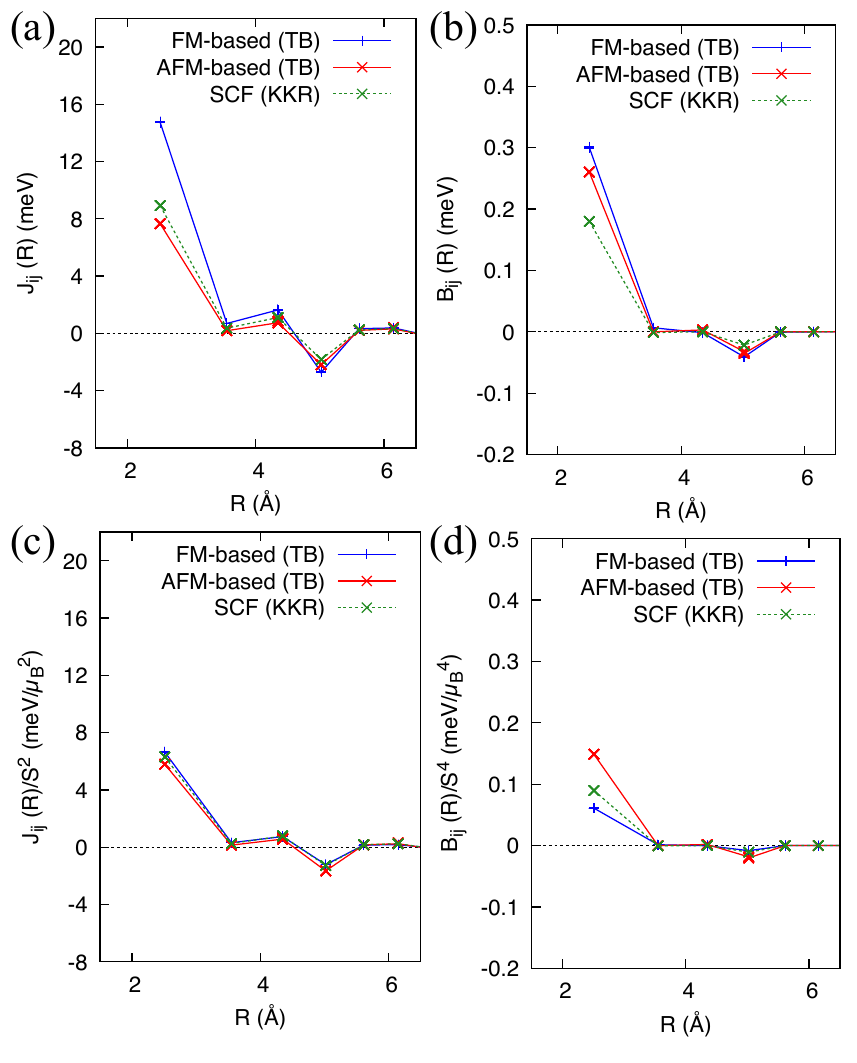}
    \caption{(a) BL, (b) BQ, (c) rescaled BL, and (d) rescaled BQ interactions of fcc Co as a function of the distance ($R$) between the $i$-th and $j$-th site. The red, blue, and green lines are those calculated using the DLM states based on the FM, the AFM state in our approach, and the KKR method, respectively.}\label{fig:cof-jij-rdep-comp-kkr}
\end{figure}

Following the discussion on hcp Co, another example of a magnet with softer magnetic moments is fcc Co.
To examine the influence of the starting magnetic order, we consider an AFM state where two magnetic moments in the $2\times{1}\times{1}$ supercell of the primitive cell are oriented in opposite directions.
In Table \ref{tb:cof-comp-momsize}, we list the magnitudes of the magnetic moments of the FM-based and AFM-based DLM states obtained using our approach, along with the SCF-DLM state.
We also show spin interactions in Fig.~\ref{fig:cof-jij-rdep-comp-kkr}.
Fig.~\ref{fig:bands} (e) and (f) show the band structures of these states from SDFT calculations and their Wannier-based tight-binding fittings.

As in the case of hcp Co, the FM-based DLM state yields magnetic moments larger than those obtained by the KKR method, resulting in stronger spin interactions in Fig.~\ref{fig:cof-jij-rdep-comp-kkr} (a) and (b).
In contrast, the AFM-based DLM state produces results that are more consistent with the SCF-DLM state calculated by the KKR method, both in terms of the magnitude of the magnetic moments and the spin interactions.
Thus, for these materials with soft magnetic moments, the parametrization starting from the AFM state to construct tight-binding parameters is the most intuitive since the state has no net magnetization.
We also present the rescaled spin interactions in panels (c) and (d).
Similar to hcp Co, the BL interactions show a good agreement.
Though the BQ interactions show some deviations, the signs and the order of magnitudes of spin interactions are common.
These results again support our findings that our approach is consistent with the SCF-DLM state when the magnetic moments are similar, and the range of the values of spin interactions is reliably estimated by the magnitude of the magnetic moments.

\begin{table}[htbp]
    \renewcommand{\arraystretch}{1.7}
    \centering
    \begin{tabular}{c|c|c|c}%{C{2.7cm}|C{1.1cm}|C{1.6cm}|C{1.6cm}}
         & KKR & FM-based & AFM-based \\ \hline\hline
        Ordered state ($\muB$) & 1.67 & 1.69 & 1.11 \\ \hline
        DLM state ($\muB$) & 1.19 & 1.49 & 1.15
    \end{tabular}
    \caption{Magnitudes of the magnetic moments ($\muB$) of the ordered states and the DLM states obtained from the FM and AFM states, and those calculated using the KKR method for fcc Co.}
    \label{tb:cof-comp-momsize}
\end{table}

Finally, we discuss the applicability and limitations of our method.
Our approach utilizes the DLM picture within the single-site CPA formalism, which accounts for transverse fluctuations of magnetic moments but neglects longitudinal ones.
Additionally, the single-site CPA considers fluctuations only locally by introducing a site-diagonal self-energy.
As a result, our method is well-suited for magnets where local magnetic moments are well-defined, but is less effective for magnets where magnetic moments are not locally polarized.

Indeed, it is well known that the DLM picture with the single-site CPA formalism fails to model the magnetism of such materials.
For instance, the SCF-DLM state calculated by the KKR method for fcc Ni shows no magnetic moment.
Extending our approach to handle such itinerant magnets would require incorporating longitudinal fluctuations~\cite{Shallcross2005-qa} or adopting the cluster-CPA formalism to capture non-local correlations~\cite{Staunton1992-et, Staunton2014}. 
Incorporating these enhancements remains an important task for future work.

%------------------------------------------------------------------------
%------------------------------------------------------------------------
%------------------------------------------------------------------------
%------------------------------------------------------------------------
\section{Applications to Complex Compounds}\label{sec:compounds}
%------------------------------------------------------------------------
%------------------------------------------------------------------------

%------------------------------------------------------------------------
%------------------------------------------------------------------------
\subsection{Co$_{1/3}$TaS$_{2}$}\label{subsec:CoTS}
This compound is an example of TMD intercalated with 3$d$ transition metal atoms.
This class of materials has been attracting extensive interest as a platform for exploring nontrivial magnetic properties~\cite{Moriya1982-dw, Checkelsky2008-ev, Mangelsen2020-gv}.
Among their intriguing properties, the anomalous Hall effect has been observed in this compound~\cite{Ghimire2018-ot}, and its origin has often been debated~\cite{Smejkal2020-cb, Park2022-ps}.
Recent studies suggest that a nontrivial magnetic structure with a four-sublattice configuration, in which each magnetic moment points toward the vertices of a tetrahedron (referred to as the All-in-All-out (AIAO) structure), is responsible~\cite{Park2023-ys, Takagi2023-uw, Dong2024-iu}.
This complex magnetic ordering has also drawn theoretical interest~\cite{Park2022-cf, Heinonen2022-hd}.

For exhibiting the AIAO structure in the ground state, the frustration of the antiferroamagnetic BL interactions between the first and second NNs in the HCP lattice formed by Co atoms is necessary at least.
Although collinear and AIAO structures are the degenerate ground states at $T=0$, the collinear structure is favored due to the order-by-disorder mechanism~\cite{Henley1989-oq, Villain1980-bm} at finite temperatures ($T\neq{0}$).
In addition to the frustration of the antiferroamagnetic BL interactions, the BQ interaction has been proposed as the key mechanism to stabilize the AIAO structure.
Indeed, finite BQ interaction has been shown to be essential for reproducing the experimental temperatures of the phase transitions of the magnetic structure in a phenomenological model~\cite{Park2023-ys}.
In this context, we apply our method to this compound and demonstrate that the calculated BQ interaction is consistent with experimental observations.  
This result highlights the applicability of our method to realistic materials for evaluating BQ interactions from first-principles.

\begin{figure}[htbp]
    \centering
    \includegraphics[width=\columnwidth, keepaspectratio]{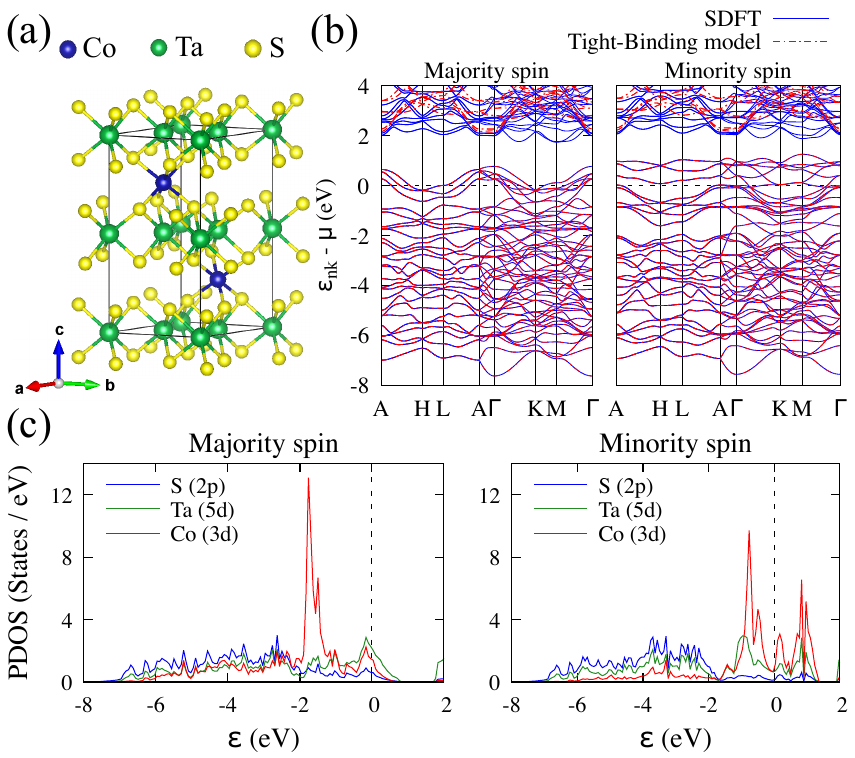}
    \caption{(a) Crystal structure of the primitive cell of Co$_{1/3}$TaS$_2$. The figure is depicted using VESTA package~\cite{Momma2011-tl}. (b) Band structure of the ferromagnetic state. The blue solid curves are the band structure calculated using SDFT, and the red dashed curves are those of the constructed tight-binding model. (c) Partial density of states of Co 3$d$, Ta 5$d$, and S 2$p$ orbitals calculated by SDFT.}
    \label{fig:CoTS-struct-band-pdos}
\end{figure}

Figure \ref{fig:CoTS-struct-band-pdos} presents (a) the crystal structure of the primitive unit cell, (b) the band structure of the FM state with that of the fitted tight-binding model for \CoTS, and (c) the partial density of states (PDOS) of Co 3$d$, Ta 5$d$, and S 2$p$ orbitals which are considered when constructing the tight-binding model.
The spin polarization of this material is primarily attributed to Co 3$d$ orbitals, while Ta 5$d$ and S 2$p$ orbitals are unpolarized and itinerant.
Although Co 3$d$ orbitals are not fully localized, as evidenced by their bands crossing the Fermi energy, the itinerant electrons mainly originate from Ta 5$d$ and S 2$p$ orbitals.
Additionally, as shown in Fig. \ref{fig:CoTS-struct-band-pdos} (a), Co atoms locate the center of the octahedra formed by S atoms and are well separated from each other, allowing them to retain locally polarized spins.
This structural and electronic configuration enables us to derive the effective classical spin Hamiltonian based on locally polarized spins on Co atoms.

\begin{figure}[htbp]
    \centering
    \includegraphics[width=\columnwidth, keepaspectratio]{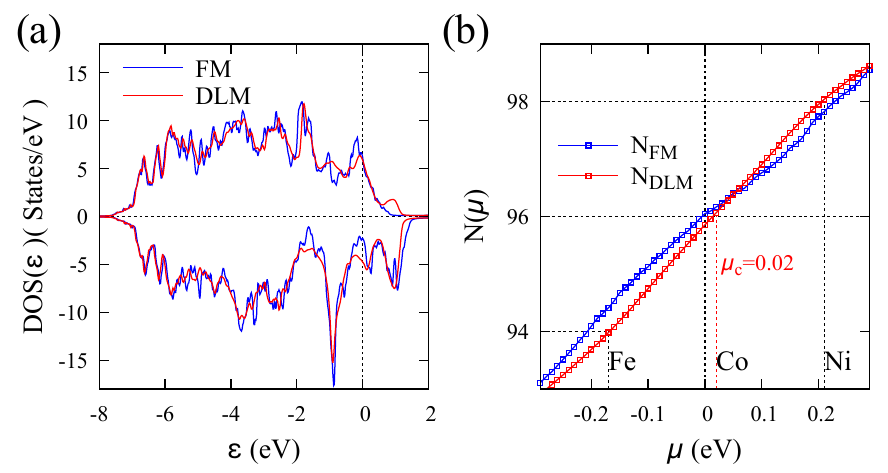}
    \caption{(a) Density of states (DOS) and (b) integrated DOS of the FM (blue) and DLM states (red). In (b), the red vertical dotted line denotes $\mu_c$. The other two black dotted lines are the chemical potential corresponding to TM$_{1/3}$TaS$_2$ (TM = Fe and Ni) from the left, respectively.}
    \label{fig:CoTS-dlm-state}
\end{figure}

We determine the DLM state for this compound based on the constructed tight-binding model.
Figure \ref{fig:CoTS-dlm-state} shows (a) the DOS and (b) the integrated DOS of the DLM state compared to those of the FM state.
Although the chemical potential of the DLM state ($\mu_c$) remains nearly unchanged from that of the FM state ($\mu=0$), the magnitude of the magnetic moment of Co atoms shrinks slightly from 1.57 $\muB$ in the FM state to 1.43 $\muB$ in the DLM state.
In Fig. \ref{fig:CoTS-dlm-state} (b), we also indicate the chemical potential corresponding to compounds intercalated with other transition metals, such as Fe and Ni.
Since the rigid-band picture is valid for this class of materials, we can qualitatively explore other compounds intercalated with different transition metals by tuning the chemical potential~\cite{Hatanaka2023-sx}.

\begin{figure}[htbp]
    \centering
    \includegraphics[width=\columnwidth, keepaspectratio]{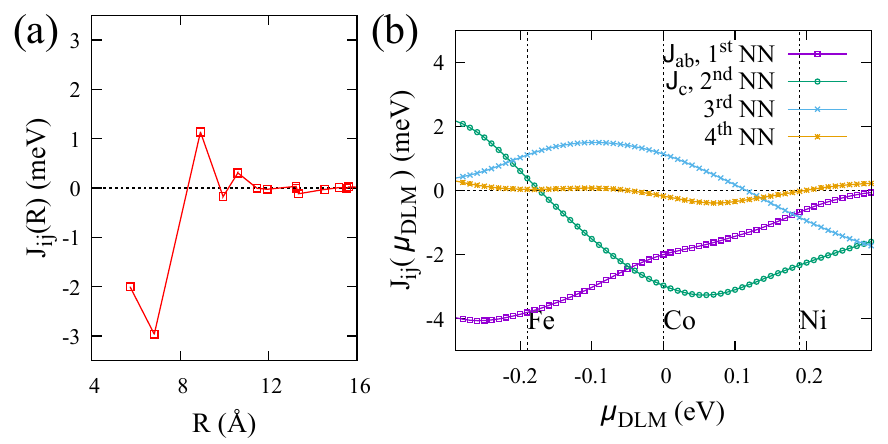}
    \caption{(a) BL spin interactions as a function of the distance between spins. (b) Chemical potential dependence of the BL interactions up to the fourth NN ones. In (b), three vertical dotted lines are the chemical potential corresponding to TM$_{1/3}$TaS$_2$ (TM = Fe, Co, and Ni) from the left, respectively.}
    \label{fig:CoTS-jij-main}
\end{figure}

\begin{figure}[htbp]
    \centering
    \includegraphics[width=\columnwidth, keepaspectratio]{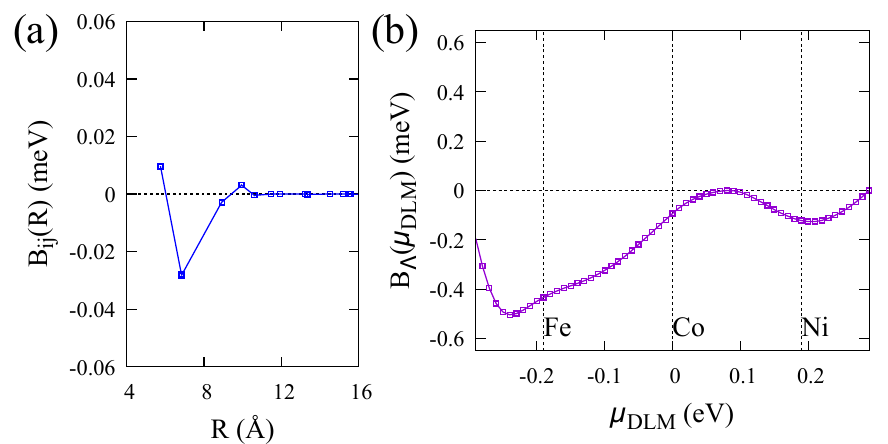}
    \caption{(a) BQ spin interactions as a function of the distance between Co spins. (b) Chemical potential dependence of $B_{\Lambda}$. In (b), three vertical dotted lines represent the chemical potential corresponding to TM$_{1/3}$TaS$_2$ (TM = Fe, Co, and Ni) from the left, respectively.}
    \label{fig:CoTS-bij-main}
\end{figure}

Then, we present the results of SCE-DLM applied to the DLM state of \CoTS in Figs. \ref{fig:CoTS-jij-main} and \ref{fig:CoTS-bij-main}.
As shown in the benchmarks, results of SCE-DLM based on \textit{ab initio} tight-binding model can depend on the starting magnetic order.
Additionally, the $U$ value in SDFT$+U$ calculations could also affect the SCE-DLM results through modifying the band structure and the tight-binding parameters.
Therefore, we explore how the value of $U$ in SDFT+$U$ calculations and the initial magnetic order affect the spin interactions in Appendices \ref{app:CoTS-jij-u} and \ref{subapp:CoTS-afm}.
Notably, the subsequent results are largely insensitive to $U$ or the initial magnetic order, as the magnitudes of the magnetic moments of Co are nearly identical in several magnetic structures, including the FM and collinear AFM structures.

In Figs. \ref{fig:CoTS-jij-main} (a) and (b), we present the BL interactions as a function of the distance between spins and the chemical potential dependence of the BL interactions up to the fourth NNs, respectively.
As shown in Fig. \ref{fig:CoTS-struct-band-pdos}, this material primarily consists of locally polarized $d$-orbitals of Co and the itinerant electrons from Ta and S atoms.
Therefore, a Kondo-type Hamiltonian can be assumed as the effective electronic Hamiltonian, and an oscillatory behavior characteristic of the RKKY interaction is observed in Fig. \ref{fig:CoTS-jij-main} (a).
In Fig. \ref{fig:CoTS-jij-main} (b), the first and second NN BL interactions, corresponding to NN interactions in the in-plane and out-of-plane directions of the $ab$-plane, are denoted as $\Jab$ and $\Jc$, respectively.
To stabilize the AIAO structure, it is necessary for both $\Jab$ and $\Jc$ to be antiferromagnetic and induce frustration, as mentioned previously.
Our result shows both $\Jab$ and $\Jc$ are antiferromagnetic, satisfying the necessary condition, and is consistent with the previous work deriving the BL interactions from the FM state~\cite{Hatanaka2023-sx}.
While \CoTS satisfies this condition, examining the chemical potentials corresponding to other TMs, i.e., Fe and Ni, reveals a tendency for either $\Jab$ or $\Jc$ to become positive, destabilizing the AIAO structure in these compounds~\cite{Mangelsen2020-gv, Fan2017-kj, An2023-rh}.
Specifically, \NiTS is observed as an antiferromagnet, in which spins are ferromagnetically aligned within the $ab$-plane and modulated along the $c$-axis.~\cite{An2023-rh}.
Since our result reveals that $J_{ab}$ tends to be positive for this compound with $J_c$ remaining negative, this result is qualitatively consistent with the experimental observation~\cite{An2023-rh}.

In Table \ref{tb:CoTS-jij-comp}, we compare the BL interactions up to the fourth NNs from our results with those of the phenomenological model in Ref.~\cite{Park2025-cd}, which were obtained by fitting experimental dynamical spin structure factor maps data of the paramagnetic state.
Therefore, the obtained interactions are independent of the magnetic ordering information and have characteristics similar to those of our SCE-DLM method.
The interactions listed in the table are scaled by the magnitude of the magnetic moment $S$.
Our results agree well with the phenomenological model in terms of the sign and order of magnitudes of the interactions.

\begin{table}[htbp]
    \renewcommand{\arraystretch}{1.7}
    \centering
    \begin{tabular}{c|cccc}%{M{2cm}|M{1.3cm}M{1.3cm}M{1.3cm}M{1.3cm}}
        $J_{ij}S^2$ (meV) & 1$^{\textrm{st}}$ $(J_{ab})$ & 2$^{\textrm{nd}}$ $(J_{c})$ & 3$^{\textrm{rd}}$ & 4$^{\textrm{th}}$ \\ \hline\hline
        SCE-DLM $(S=1)$ & -2.02 & -2.97 & 1.11 & -0.18\\ \hline
        Ref.~\cite{Park2025-cd} $(S=3/2)$ & -2.73 & -3.16 & 0.59 & -0.72
    \end{tabular}
    \caption{Summary of the BL interactions up to the fourth NNs of our result and those of the phenomenological model in Ref.~\cite{Park2025-cd}.}\label{tb:CoTS-jij-comp}
\end{table}

Subsequently, we show the SCE-DLM results for the BQ interaction.
Fig.~\ref{fig:CoTS-bij-main} presents (a) the BQ interaction as a function of the distance between spins and (b) the chemical potential dependence of the sum of the BQ interactions that lift the degeneracy with the collinear state.
$(\bm{S}_{i}\cdot\bm{S}_{j})^2$ takes the values 1 or 1/9 in the AIAO structure.
The former coincides with that of the collinear structure, where the BQ interactions do not lift the degeneracy.
In contrast, in the latter case, the BQ interactions lift the degeneracy.
We hereafter define $\Lambda$ as $\Lambda=\{j~|~(\bm{S}_{0}\cdot\bm{S}_{j})^2=1/9\}$ in the AIAO structure.
Specifically, $\Lambda$ includes the first, second, and fourth NN interactions from the origin site, with more distant interactions being negligible.
Our analysis focuses on the sum of the BQ interactions between the sites in $\Lambda$.
Therefore, we define $B_{\Lambda}$ as the summed value, i.e., $B_{\Lambda} = \sum_{j\in\Lambda}B_{0j}$.
As investigated in Appendix~\ref{app:CoTS-jij-u}, while individual BQ interactions are sensitive to the $U$ value in SDFT$+U$ calculations, both in sign and magnitude, $B_{\Lambda}$ remains robust against variations in $U$.

To stabilize the AIAO structure, $B_{\Lambda}$ is required to be negative, in addition to the frustration caused by the BL interactions.
We observe that $B_{\Lambda}$ is negative at $\muDLM=0$, indicating that \CoTS satisfies this condition.
For \TMTS (TM = Fe and Ni), the BL interactions tend to fail to meet the necessary conditions to stabilize the AIAO structure.
Consequently, the AIAO structure cannot be stabilized regardless of the sign of $B_{\Lambda}$ in these materials.
Although \CoTS~satisfies the condition at $\muDLM = 0$, an increase in $\muDLM$ tends to make $B_{\Lambda}$ positive.
This suggests that the AIAO structure is fragile against electron doping, which is consistent with experimental observations~\cite{Park2024-hz}.
In the experiment, when the composition ratio $x$ of Co atoms gradually increases, the AIAO structure is no longer the magnetic structure of the ground state around $x\sim{0.33}$.
We find that the chemical potential dependence of $B_{\Lambda}$ qualitatively reproduces the experimental trends.
This high sensitivity to the chemical potential in stabilizing non-coplanar structures, such as the AIAO structure, is also observed in theoretical models.
For the Kondo Hamiltonian on the triangular lattice, chiral magnetic ordering with finite scalar spin chirality emerges only around 1/4 or 3/4 filling~\cite{Martin2008-sk, Akagi2010-xy}.

Regarding the magnitude of the BQ interaction, it has been shown that $B/J$, where $J$ and $B$ are the NN BL and BQ interactions, respectively, must be less than approximately 0.04 to exhibit the two-step phase transition of the magnetic structure~\cite{Park2023-ys}.
Based on the BL interactions in the lower row of Table \ref{tb:CoTS-jij-comp}, $B/J$ is estimated at 0.027 to reproduce the experimental transition temperature~\cite{Park2025-cd}.
Our calculations yield a value of approximately 0.01, which is small enough to exhibit the two-step phase transition.
These results indicate that the BQ interactions calculated by our method capture the features of this compound, i.e., the stabilization of the AIAO structure, the two-step magnetic-structure transition, and the strong dependence of the stability of the AIAO state on the chemical potential.
Recently, it has been pointed out that, though the BQ interaction is sufficient to model the most general classical ground state, the four-spin interaction and magnetic anisotropy could also play a pivotal role in this compound~\cite{Kirstein2025-sb}.
The construction of a complete spin model including fourth-order spin interactions and magnetic anisotropies is left for future work.

%------------------------------------------------------------------------
%------------------------------------------------------------------------
\subsection{Potassium Electrosodalite}\label{subsec:PES}
\begin{figure}[htbp]
    \centering
    \includegraphics[width=\columnwidth, keepaspectratio]{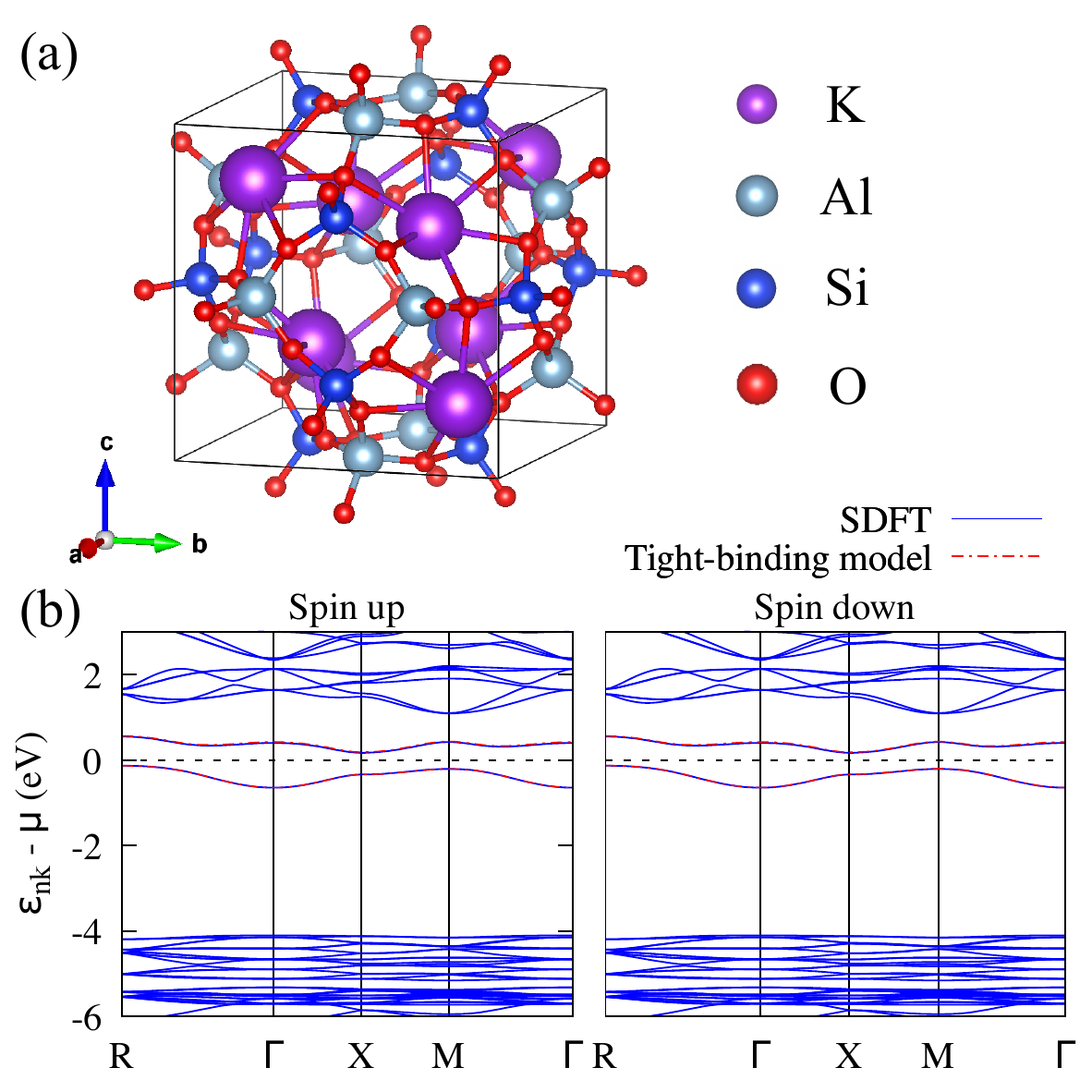}
    \caption{(a) Crystal structure of the primitive cell of \PES. The figure is depicted using the VESTA package. (b) Band structure of the antiferromagnetic state. The blue solid curves are the band structure calculated using SDFT, and the red dashed curves are those of the constructed tight-binding model.}
    \label{fig:PES-structure-band}
\end{figure}

Unlike previous applications where our method was used for systems with individual atoms hosting magnetic moments, it is also applicable to larger systems exhibiting magnetic moments in interstitial regions.
We therefore apply our method to zeolite systems known for their multifunctionality and widely studied as platforms for investigating many-body electron correlation effects~\cite{Auerbach2003-xi, Arita2004-kk}.
\PES~(Potassium electrosodalite) is one such system, which exhibits robust antiferromagnetic order~\cite{Tou2001-jm, Igarashi2010-cn} despite being composed entirely of non-magnetic elements.
Although the material comprises a large number of atoms, first-principles calculations have revealed that its low-energy electronic structure near the Fermi level is remarkably simple.
In Fig. \ref{fig:PES-structure-band}, we show (a) the crystal structure and (b) the band structure of the AFM state, from which we construct the tight-binding model.
As shown in Fig. \ref{fig:PES-structure-band}~(a), this material consists of the aluminosilicate cages that trap one valence $s$ electron per cage.
Therefore, the interstitial $s$ orbitals are located at the origin and the center of the unit cell.
As previously discussed in Ref. ~\cite{Nakamura2009-th}, owing to this property, the system can be regarded as the BCC lattice of hydrogen-like $s$ orbitals in the interstitial region.
Indeed, as shown in Fig. \ref{fig:PES-structure-band}~(b), the band structure of the antiferromagnetic state shows only two bands around the Fermi level and exhibits a gap, indicating an antiferromagnetic insulator.

In Ref.~\cite{Nakamura2009-th}, the Hubbard model was derived using the constrained random-phase approximation (cRPA) method~\cite{Aryasetiawan2006-mm}, from which the Heisenberg model was subsequently constructed using strong coupling expansion. Here, we apply our method to this system and demonstrate that the calculated exchange interactions show comparable agreement with experimental data compared to the cRPA approach~\cite{Nakamura2009-th}.
These results indicate that our method is effective for systems with a large number of atoms by downfolding to a minimal tight-binding effective model.

\begin{figure}[htbp]
    \centering
    \includegraphics[width=\columnwidth, keepaspectratio]{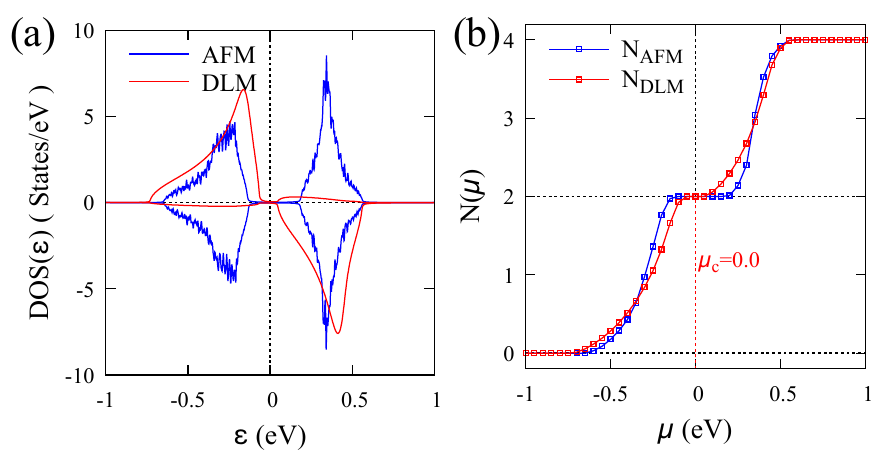}
    \caption{(a) Density of states (DOS) and (b) integrated DOS of the antiferromagnetic (blue) and DLM states (red). In (b), the red vertical dotted line denotes $\mu_c$.}
    \label{fig:PES-dlm-state}
\end{figure}

We determine the DLM state from the obtained tight-binding model.
Figure \ref{fig:PES-dlm-state} shows (a) the DOS and (b) the integrated DOS of the DLM state compared to those of the AFM state.
From Fig.~\ref{fig:PES-dlm-state}(a), it is evident that, despite the broadening of the electronic structure caused by thermal disorder~\cite{Bouaziz2025-bl}, the DLM state remains insulating, similar to the AFM state.
Additionally, both the chemical potential ($\mu_c$) and the magnitude of the magnetic moment of the interstitial $s$ orbitals remain unchanged from those of the AFM state.
Specifically, the magnitudes of the magnetic moments in the AFM and the DLM states are 0.88 and 0.89~$\muB$, respectively.
In appendix \ref{subapp:PES-afm}, we compare subsequent results with those obtained from the FM state.
The results are almost identical to subsequent results based on the AFM state.

\begin{figure}[htbp]
    \centering
    \includegraphics[width=\columnwidth, keepaspectratio]{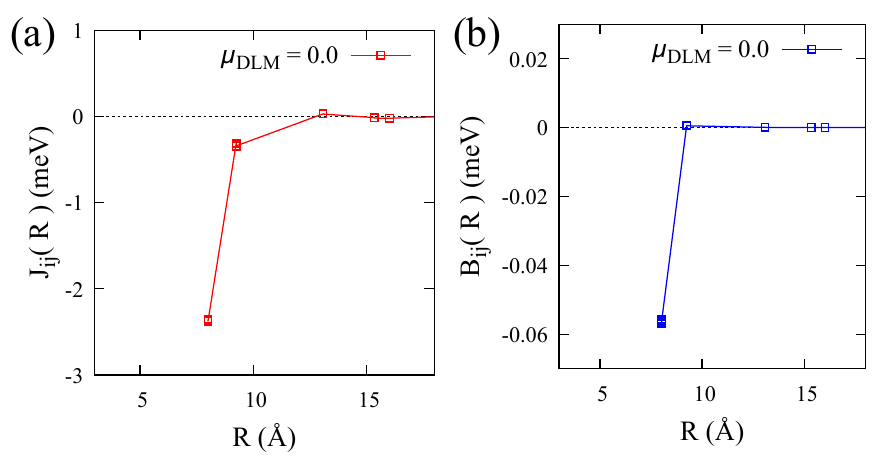}
    \caption{
    (a) BL and (b) BQ spin interactions for \PES, calculated as a function of the distance between the spins polarized in the interstitial regions.
    }
    \label{fig:PES-jij-bij}
\end{figure}

Fig.~\ref{fig:PES-jij-bij} shows (a) the BL and (b) BQ interactions as a function of the distance.
Since the DLM state remains insulating, the first and second NN interactions are dominant, while the other distant interactions are negligibly small.
In Table~\ref{tb:PES-jij-comp}, we summarize the BL interactions of our work and those in the previous theoretical work~\cite{Nakamura2009-th}.
In addition to that, we present the experimental value obtained by fitting the experimental N\'{e}el and Weiss temperature~\cite{Nakano2010-kc} by the Heisenberg model defined in Ref.~\cite{Nakamura2009-th}, consisting of only the first and second NN interactions.
Although our BL interactions are larger than the experimental values, their agreement with experiment is comparable to that of the previous theoretical result.
Furthermore, the BQ interactions obtained by our method are negative and favor non-collinear structures.
As a result, the BQ interactions effectively weaken the BL interaction, making the BL interaction much closer to the experimental values.
Our approach is applicable to compounds containing a large number of atoms and hosting magnetic moments even in interstitial regions without atomic sites through mapping onto Wannier orbitals.
In contrast, the methods based on KKR or LMTO are generally not suitable for this type of compound.
This highlights the versatility of our method based on the tight-binding model.

\begin{table}[htbp]
    \renewcommand{\arraystretch}{1.5}
    \centering
    \begin{tabular}{c|cc}%{M{6cm}|M{1.3cm}M{1.3cm}} %{c|cc}
        $J_{ij}S^2$ (meV) & 1$^{\textrm{st}}$ & 2$^{\textrm{nd}}$ \\ \hline\hline
        SCE-DLM $(S=1)$ & -2.36 & -0.33 \\ \hline
        cRPA and strong coupling expansion~\cite{Nakamura2009-th} $(S=1/2)$ & -0.30 & -0.04\\ \hline
        Expt.~\cite{Nakamura2009-th, Nakano2010-kc} $(S=1/2)$& -1.03 & -0.43
    \end{tabular}
    \caption{Summary of the first and second NN BL interactions of our result and those obtained from the cRPA and strong coupling expansion in Ref.~\cite{Nakamura2009-th}. The experimental values were obtained by fitting the experimental N\'{e}el and Weiss temperatures~\cite{Nakano2010-kc} in Ref.~\cite{Nakamura2009-th}.}\label{tb:PES-jij-comp}
\end{table}

%------------------------------------------------------------------------
%------------------------------------------------------------------------
%------------------------------------------------------------------------
%------------------------------------------------------------------------
\section{Conclusion}\label{sec:conclusion}
We developed the SCE-DLM approach for \textit{ab initio} tight-binding Hamiltonians by combining the spin cluster expansion and the disordered local moment (DLM) approach, which was originally available only with the Green's function methods, such as the KKR method.
We first applied the SCE-DLM scheme to the one-dimensional Hubbard model with two sublattices and found that the calculated biquadratic interactions closely aligned with those for the effective quantum spin model in the strongly correlated limit super-exchange.
This alignment suggests the broad applicability of the SCE-DLM approach to a wide variety of strongly correlated compounds with large local magnetic moments.
We subsequently applied this scheme to elemental magnetic metals, i.e., bcc Fe, hcp Co, and fcc Co.
Our results are consistent with previous theoretical investigations, confirming the reliability of the presented method.
In addition, we corroborated our \textit{ab initio} tight-binding approach by comparing our results with those obtained via the KKR multiple scattering theory and showing a good agreement between these two distinct methods depending on the starting magnetic order.
In the case of bcc Fe, where local magnetic moments are well-defined, the FM state serves as a reliable starting magnetic order.  
Conversely, for hcp and fcc Co, which exhibit softer magnetic moments, the AFM state is the most intuitive, as the state has no net magnetization and provides results that are more consistent with the KKR method.
From these materials, we demonstrate that our approach produces nearly identical spin interactions when the magnitude of the magnetic moments is quantitatively comparable to that obtained via the KKR method.  
Furthermore, we show that even when different starting magnetic structures are used, spin interactions can be estimated through rescaling based on the magnitudes of the magnetic moments.
To extend this approach to more itinerant magnets, it is necessary to incorporate longitudinal fluctuations and nonlocal correlations of magnetic moments.

To demonstrate the effectiveness and general applicability of our method, we further applied it to two complex compounds: \CoTS~ and \PES.
For \CoTS, our method yielded both bilinear and biquadratic interactions that are in good agreement with experimental results.
Not only the sign and magnitude, but also the chemical potential dependence of the interactions was consistent with experiments.  
For \PES, we showed that an effective spin model can be derived at a low computational cost, even for structurally complex compounds where magnetic moments emerge in interstitial regions.
The resulting model exhibits comparable agreement with experimental results compared to other theoretical approaches.  
Furthermore, for both compounds, the dependence on the starting magnetic order was found to be sufficiently weak.  
These results highlight that our method works well for realistic magnetic compounds, where local magnetic moments can be regarded as well-defined.

The SCE-DLM scheme for the \textit{ab initio} tight-binding Hamiltonian has the advantage of minimizing the reference state dependence in the magnetic interactions, except through the tight-binding parameters.
Additionally, it requires only tight-binding parameters that can be computed based on different \textit{ab initio} schemes.
The extension of our method to incorporate relativistic chiral (Dzyaloshinskii-Moriya like) and multi-spin / multi-site interactions offers a promising approach to systematically compute the magnetic interactions between local moments to arbitrary order.
It will be a convenient tool for understanding and predicting nontrivial exotic magnetic phases induced by higher-order and relativistic spin interactions in an \textit{ab initio} framework.

%------------------------------------------------------------------------
%------------------------------------------------------------------------
%------------------------------------------------------------------------
%------------------------------------------------------------------------
\section*{Acknowledgements}
We would like to thank Katsuhiro Tanaka and Hiroshi Katsumoto for fruitful discussions. This work was supported by RIKEN Junior Research Associate Program. This work was supported by the RIKEN TRIP initiative (RIKEN Quantum, Advanced General Intelligence for Science Program, Many-body Electron Systems). We acknowledge the financial support by Grant-in-Aids for Scientific Research (JSPS KAKENHI) Grant Numbers JP21H04437, JP21H04990, JP19H05825, JP22H00290, JP24K00581, JP25H00420, JP25H01252, JP25K21684, and JP25KJ1169, JST-CREST No.~JPMJCR18T3, No.~JPMJCR23O4, JST-ASPIRE No.~JPMJAP2317, JST-Mirai No.~JPMJMI20A1.
T. N. was supported by JST, PRESTO Grant Number JPMJPR20L7, Japan. J. B. was supported by the Alexander von Humboldt Foundation through the Feodor Lynen Research Fellowship for Postdocs.

%------------------------------------------------------------------------
%------------------------------------------------------------------------
%------------------------------------------------------------------------
%------------------------------------------------------------------------
\appendix
\makeatletter
\@addtoreset{figure}{section}
\makeatother
\renewcommand{\thefigure}{\Alph{section}\arabic{figure}}
\setcounter{figure}{0}

%------------------------------------------------------------------------
\section{Asymptotic Form for Single-orbital Tight-binding Model}\label{app:asymp-sorb}
% ----------------------------------------------------------------------

In Section \ref{subsec:2-orb}, a tight-binding Hamiltonian with a spin splitting was derived from the two-sublattice Hubbard Hamiltonian. Here, let us look into a simpler case, i.e., a one-dimensional single-orbital and single-sublattice model with the mean-field approximation:
\begin{align}\label{eq:tbham-sorb}
    \mathcal{H} =& -\sum_{\langle{i,j}\rangle,\sigma}(t\,c^{\dagger}_{i\sigma}c_{j\sigma}+\mathrm{h.c.}) - B\sigma_z.
\end{align}

We can analytically obtain the on-site component of the Green's function of the DLM state for this model as follows:
\begin{align}
    \gbar_{ii}(\epsilon) &= \left\{\textrm{sgn}\qty(\Re(\epsilon-\tilde{\Sigma}^i))\sqrt{{(\epsilon-\tilde{\Sigma}^i)}^2-4t^2}\right\}^{-1}.
\end{align}
By substituting this expression of $\gbar_{ii}$ to Eqs. (\ref{eq:def-t-matrix}) and (\ref{eq:cpa-condition}) with $V(\pm\hat{\bm{z}})=\mp{B}$ for the up and down spins, the CPA condition yields the equation for the self-energy $\Sigma$ as follows:
\begin{align}\label{eq:cpa-sorb}
    0 &= 2\epsilon\tilde{\Sigma}^{3} - (2B^{2}-4t^2+\epsilon^{2})\tilde{\Sigma}^{2} + B^{4}
\end{align}

Let us now consider deriving the asymptotic expression of the exchange interaction for the limit of strong and weak correlation based on SCE-DLM.
Starting with Eq. (\ref{eq:jijl}), we expand it as follows:
\begin{align}
    J_{ij}^{LL'}\sim&\,\frac{1}{\pi}\Im \int\dd\epsilon{f}(\epsilon) \iint\dd^2\bme_i\dd^2\bme_j Y_L(\bme_i)Y_{L'}(\bme_j)\nonumber\\
    &\,\times\Big[T_{i}(\bm{e}_i)\gbar_{ij}T_{j}(\bm{e}_j)\gbar_{ji}+\nonumber\\
    &\,\quad\frac{1}{2}{T}_{i}(\bm{e}_i)\gbar_{ij}T_{j}(\bm{e}_j)\gbar_{ji}T_{i}(\bm{e}_i)\gbar_{ij}T_{j}(\bm{e}_j)\gbar_{ji}\Big],
    \label{eq:jijl-expand}
\end{align}
Here, we utilize the Taylor expansion, $\log(1-x)\sim -x-x^2/2$.
Though there are other higher-order terms in the expansion, these terms are sufficient to obtain leading-order terms of the bilinear (BL) and biquadratic (BQ) interactions in the strong and itinerant limits.
We also derive the expression of the scattering operator $T_i(\bme_i)$ by applying the CPA condition in Eq.(\ref{eq:cpa-sorb}) and introducing the inverse of the Green's function as $A=1/\gbar_{ii}(\epsilon)$,
\begin{align}\label{eq:ti-ei}
    T_{i}(\bme_i) = -\frac{BA^2}{{(A+\tilde{\Sigma})}^2-B^2}\mqty(\cos\theta&e^{-i\phi}\sin\theta\\e^{i\phi}\sin\theta&-\cos\theta).
\end{align}

\subsection{Itinerant Limit}
In the limit of $B\ll{t}$, the CPA condition becomes $0 = 2\epsilon\tilde{\Sigma}^{3} + (4t^2-\epsilon^{2})\tilde{\Sigma}^{2}$ and the solutions of this equation are:
\begin{align}
    \tilde{\Sigma}(\epsilon) &= 0,\, \frac{\epsilon}{2}+\frac{2t^2}{\epsilon}
\end{align}
As the latter solution is unphysical in the limit of $\epsilon\rightarrow\pm\infty$, the solution of the CPA condition approaches $\tilde{\Sigma}(\epsilon) \rightarrow 0$.
By substituting this self-energy solution to Eq. (\ref{eq:ti-ei}), the scattering operator becomes
\begin{align}
    T_{i}(\bme_i) \rightarrow -B\mqty(\cos\theta&e^{-i\phi}\sin\theta\\e^{i\phi}\sin\theta&-\cos\theta).
\end{align}
We here use that $A\rightarrow\sqrt{\epsilon^2-4t^2}$.
The first term of the right-hand side in Eq. (\ref{eq:jijl-expand}) remains finite only for $l=1$.
This corresponds to the fact that higher-order interactions ($l\ge{2}$) require perturbations higher than the second order.
It is also important to note again that $J^{(l,m)(l,m)}_{ij}$ depends solely on $l$ and is independent of $m$ in the absence of SOC.
Then, we can easily show that the BL interaction in SCE-DLM (Eq.~(\ref{eq:jijl-expand})) becomes equivalent to that of LKAG in this limit, hence yielding the RKKY interaction~\cite{ruderman, kasuya, yoshida} as follows:
\begin{align}
    J_{ij} &= \frac{3}{8\pi}J_{ij}^{(1,0)(1,0)}\nonumber\\
    &\rightarrow \frac{B^2}{\pi{N}^2}\sum_{k,q}\Im\int\dd\epsilon f(\epsilon)G^0_{k+q}G^0_{k}e^{iqR_{ij}}\nonumber\\
    &= \frac{B^2}{\pi{N}}\sum_{q}\chi(q)e^{iq{R_{ij}}},\label{eq:asymp-j-itinerant}
\end{align}
where $N, G^0_{k}=(\epsilon-\epsilon_{k}+i\delta)^{-1}, \chi(q)$ is the number of sites, the retarded Green's function, and the spin susceptibility of non-interacting electrons, respectively.
Here we use an infinitesimally small value $\delta$, and $\chi(q)$ is defined as follows:
\begin{align}
    \chi(q) &= \frac{1}{\pi{N}}\sum_{k}\Im\int\dd\epsilon f(\epsilon)G^0_{k+q}G^0_{k}\\
    &= \frac{1}{N}\sum_{k}\frac{f(\epsilon_{k})-f(\epsilon_{k+q})}{\epsilon_{k+q}-\epsilon_{k}+i\delta}.
\end{align}

With SCE-DLM, we can obtain higher-order interactions such as the BQ interaction.
Indeed, we can derive the BQ interaction by considering higher-order terms in Eq.~(\ref{eq:jijl-expand}).
The asymptotic expression of the BQ interaction becomes:
\begin{align}
    &B_{ij} = \frac{15}{16\pi}J_{ij}^{(2,0)(2,0)}\nonumber\\
    &\rightarrow \frac{B^4}{\pi{N}^2}\sum_{k,q,k',q'}\Im\int\dd\epsilon f(\epsilon)G^{0}_{k+q}G^0_{k}G^{0}_{k'+q'}G^0_{k'}e^{i(q+q'){R_{ij}}}\label{eq:asymp-b-itinerant}.
\end{align}
As discussed in Ref. \cite{eff-blbq-hym}, contributions from the $k=k', q=q'$ case always yield a negative BQ interaction between any sites and cause instability of the collinear spin structures.

\subsection{Strongly Correlated Limit}
In the limit of strong correlation $U\sim{B}\gg{t}$, the CPA condition becomes to $2\epsilon\tilde{\Sigma}^3-(2B^2+\epsilon^2)\tilde{\Sigma}^2 + B^4 = 0$ and the solutions are given as: 
\begin{align}
    \tilde{\Sigma}(\epsilon) = \frac{B^2}{\epsilon},\, \frac{\epsilon\pm\sqrt{\epsilon^2+8B^2}}{4}.
\end{align}
Similarly to the itinerant case, the solution of the self-energy is $\tilde{\Sigma}(\epsilon) \rightarrow B^2/\epsilon$.

We start from the DLM state without the hopping term $t$. The retarded Green's function for this non-perturbed state is provided as:
\begin{align}
    \gbar^{(0)}_{ii}(\epsilon) = \frac{\delta_{ij}}{\epsilon-\Sigma+i\delta},
\end{align}
where the superscript $(0)$ stands for the non-perturbed term.
By treating the hopping term as the perturbation, we can derive the expression for the Green's function, considering terms up to the first order perturbation:
\begin{align}
    \gbar^{(1)}_{ij}(\epsilon) &= \frac{1}{\epsilon-\tilde{\Sigma}+i\delta}t\frac{1}{\epsilon-\tilde{\Sigma}+i\delta}\nonumber\\
    &=\frac{t}{{(\epsilon-\tilde{\Sigma}+i\delta)}^2}\label{eq:gt1},
\end{align}
where the superscript $(1)$ stands for the first order perturbation term and $j$ is the nearest-neighbor sites of site $i$.

\begin{figure}[htbp]
    \centering
    \includegraphics[keepaspectratio, width=0.6\columnwidth]{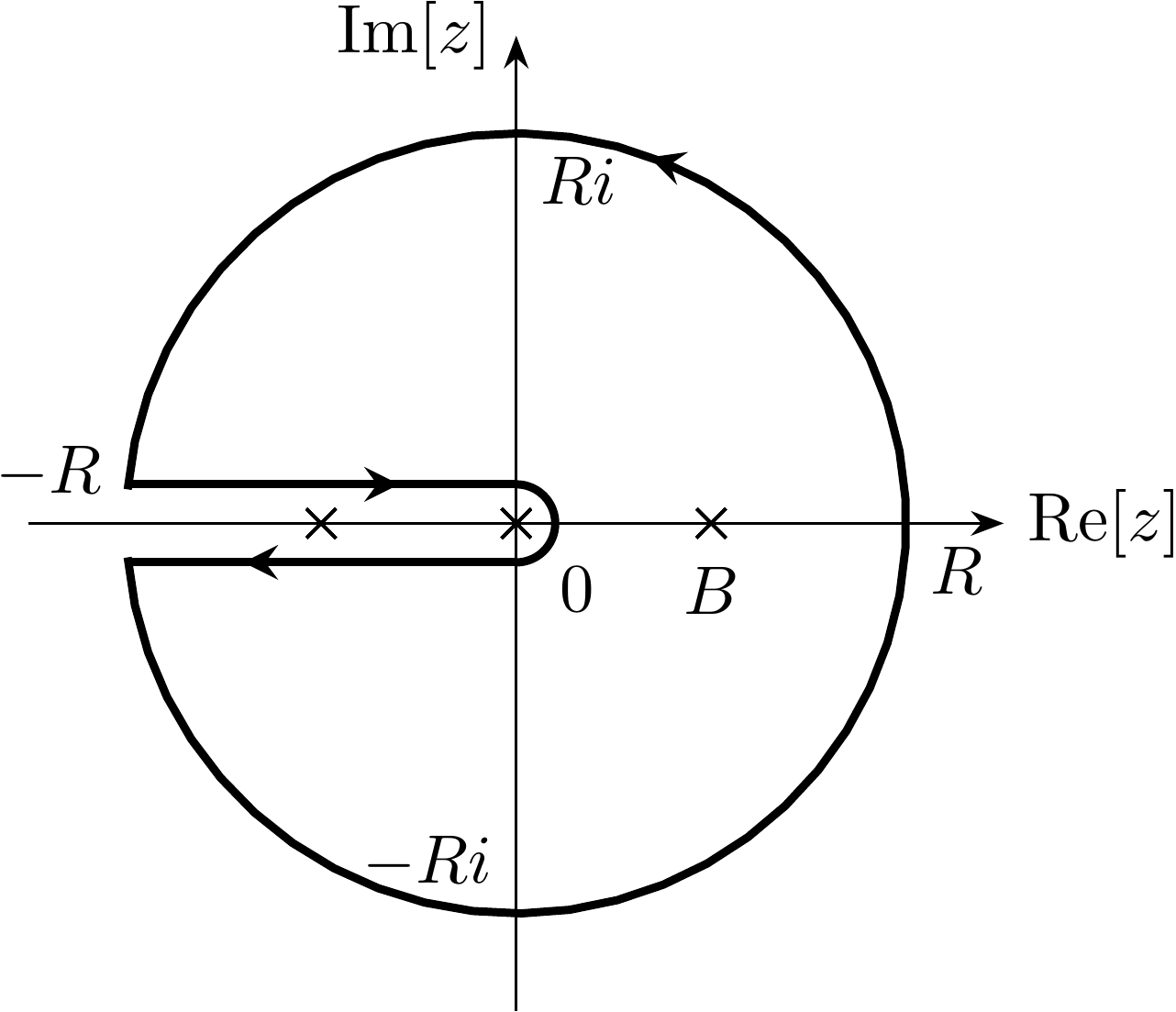}
    \caption{Integration contour of Eq. (\ref{eq:int-zfunc}). We take the $R\rightarrow\infty$ limit in the integration.}\label{fig:contour}
\end{figure}

Subsequently, we evaluate the scattering operator as follows:
\begin{align}
    T_{i}(\bm{e}_i) \rightarrow&-\frac{{(\epsilon-\tilde{\Sigma})}^2}{\epsilon^2-B^2}B\mqty(\cos\theta&e^{-i\phi}\sin\theta\\e^{i\phi}\sin\theta&-\cos\theta),
    \label{eq:asymp-te}
\end{align}
where we use $A\rightarrow{\epsilon-\tilde{\Sigma}}$ in this limit.
By substituting Eqs. (\ref{eq:gt1}) and (\ref{eq:asymp-te}) into the first term of Eq. (\ref{eq:jijl-expand}), we obtain the following asymptotic expression for the $l=1$ interaction between nearest-neighbor sites:
\begin{align}
    J_{ij}^{(1,m)(1,m)} \sim&\,-\frac{2}{\pi}\frac{4\pi}{3}\Im \int^{\epsilon_F}\dd{\epsilon}\frac{t^2B^2}{{(\epsilon-\Sigma+i\delta)}^4}\frac{{(\epsilon-\Sigma)}^4}{{(\epsilon^2-B^2)}^2}\nonumber\\
    =&\,-\frac{4}{3i}\int_{C}\dd{z}\frac{t^2B^2}{{(z-B)}^2{(z+B)}^2}.\label{eq:int-zfunc}
\end{align}
We illustrate an integration contour in Eq. (\ref{eq:int-zfunc}) in Fig. \ref{fig:contour}.

We subsequently derive an asymptotic expression for the BL interaction.
\begin{align}\label{eq:asymp-j-correlatred}
    J_{ij} &= \frac{3}{8\pi}J_{ij}^{(1,0)(1,0)}\rightarrow -\frac{t^2}{4B}\sim-\frac{t^2}{2U}
\end{align}
We can obtain the expression for the BQ interaction by following the same process.
\begin{align}\label{eq:asymp-b-correlatred}
    B_{ij} =& \frac{15}{16\pi}J_{ij}^{(2,0)(2,0)}\rightarrow-\frac{5}{4}\frac{t^4}{U^3}.
\end{align}

These asymptotic expressions for the BL and BQ interactions are equivalent to those obtained by the conventional LKAG method and its extensions\cite{extension} in both the strongly correlated and the itinerant limits.
However, it is crucial to recognize that the initial magnetic state in SCE-DLM, the DLM state, differs from the ferromagnetic state used in these methods.
Furthermore, it is noteworthy that when using SCE-DLM and the method described in Ref. \cite{extension}, the BQ interaction remains finite even in a system consisting of only a spin-1/2 degree of freedom, in contrast to the effective quantum spin model where this term inevitably vanishes. 
In the quantum spin model of an $S=1/2$ system, this term, corresponding to a fourth-order perturbation, is merely a correction to the BL interaction.
However, the classical treatment of spins in electron systems ensures that these higher-order interactions remain finite even in a single-orbital system.
Therefore, it is not appropriate to simply compare this expression with Eq. (\ref{eq:jij-pert}).

% ----------------------------------------------------------------------
\section{Comparison with LKAG Method in a Single-orbital Tight-binding Model}\label{app:comp-lkag}
% ----------------------------------------------------------------------

To assess the sensitivity to the starting magnetic order, we compare the SCE–DLM approach with the LKAG method for a simple model: a single-orbital and single-sublattice model on the square lattice with the mean-field approximation as in Appendix \ref{app:asymp-sorb}.
We set the NN hopping $t=1$ and exchange splitting $B=3$, respectively. In the SCE-DLM calculation, the inverse temperature $\beta$ was set to 500 eV$^{-1}$, and a $64\times{64}\times{1}$ $k$-point grid was used.

\begin{figure}[htbp]
    \centering
    \includegraphics[keepaspectratio,width=0.7\columnwidth,page=1]{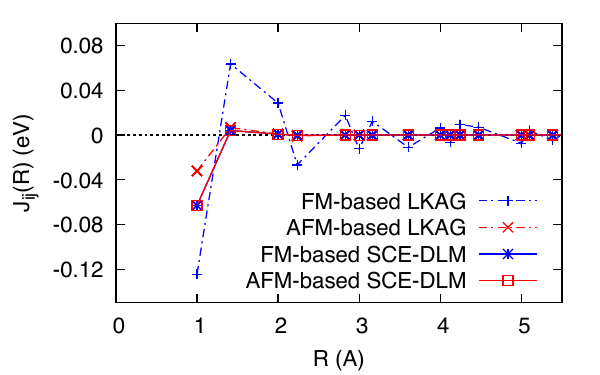}
    \caption{BL interactions versus distance in the single-orbital square-lattice tight-binding model ($t=1, B=3$, half-filling) for four cases: FM-based LKAG, AFM-based LKAG, FM-based SCE-DLM, and AFM-based SCE-DLM.}\label{fig:comp-lkag}
\end{figure}

We performed both methods starting from the FM and AFM ($\bm{q}=(\pi,\pi)$) reference states. Hereafter, we refer to the four results as FM-based LKAG, AFM-based LKAG, FM-based SCE–DLM, and AFM-based SCE–DLM. Fig. \ref{fig:comp-lkag} shows the BL exchange interactions at half-filling as a function of distance for all four cases.

The SCE–DLM results are nearly identical for the two reference states, whereas the LKAG results show a noticeable difference.
In this model, the FM state is metallic and the AFM state is insulating at half-filling.
Within LKAG, this disparity in the underlying electronic structure directly carries over to the extracted spin interactions.
By contrast, in SCE–DLM, the FM- and AFM-based DLM states are both insulating at half-filling and yield almost the same BL interactions.
This demonstrates that SCE–DLM is less sensitive to the choice of reference state than LKAG.

% ----------------------------------------------------------------------
\section{SDFT+$U$ Calculations for \CoTS}\label{app:CoTS-sdft-u}
% ----------------------------------------------------------------------

\begin{figure*}[htbp]
    \centering
    \includegraphics[width=1.9\columnwidth, keepaspectratio]{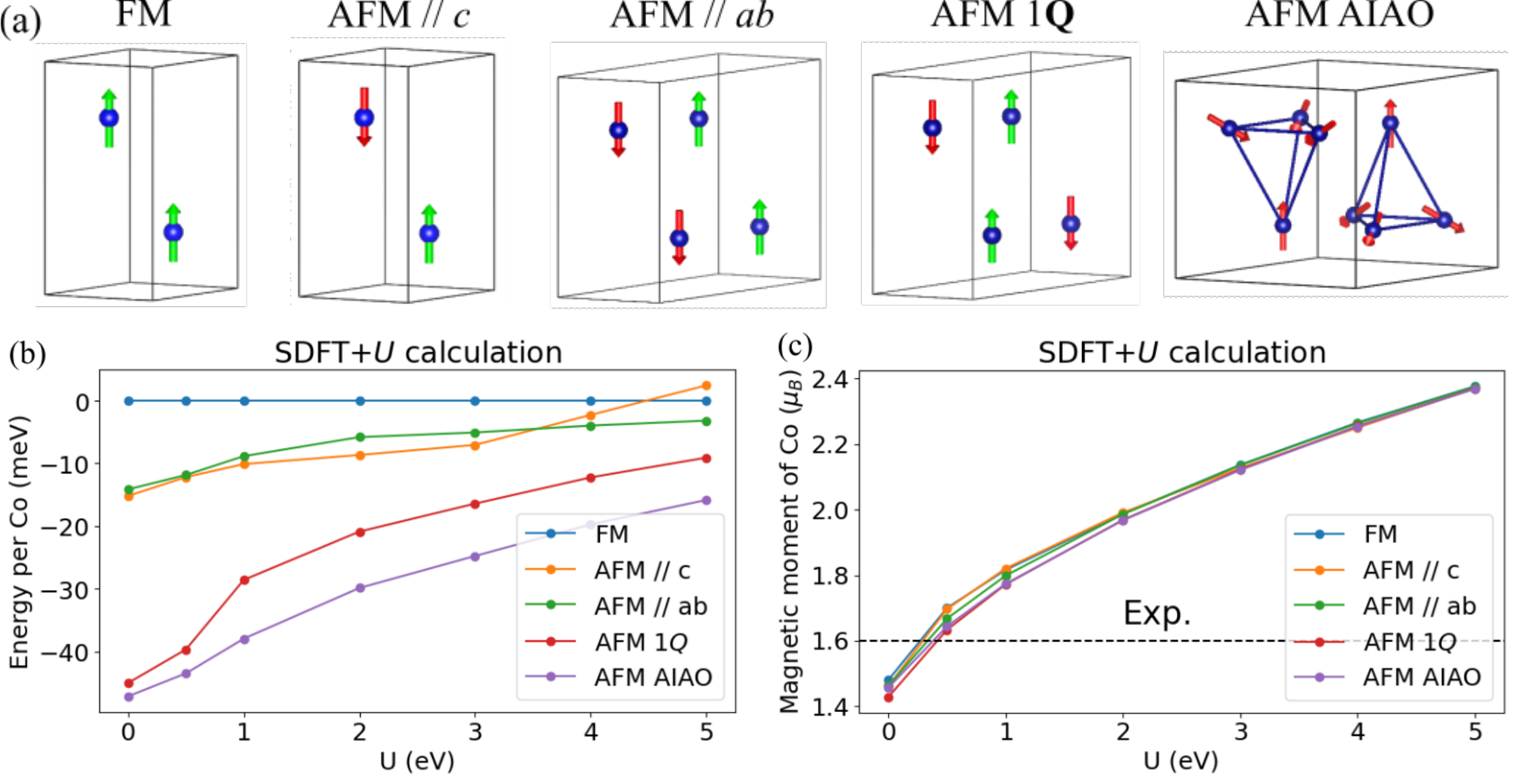}
    \caption{(a) Magnetic structures considered in the SDFT$+U$ calculations. In the magnetic structures other than the AFM AIAO structure, the light-green and red arrows represent up and down spins with respect to the quantization axis, respectively. (b) SDFT$+U$ calculation results of the energy per Co atom. (c) SDFT$+U$ calculation results of the magnitude of the magnetic moment of Co atoms, with the dashed horizontal line indicating the experimental value~\cite{Takagi2023-uw}.}
    \label{fig:CoTS-sdft-u}
\end{figure*}

We performed SDFT$+U$ calculations and determined the value of $U$ to match the magnitude of the magnetic moment of Co atoms with the experimental result.
Figure \ref{fig:CoTS-sdft-u} summarizes our SDFT+$U$ calculations. In Fig. \ref{fig:CoTS-sdft-u} (a), there are magnetic structures considered in these calculations.
Only magnetic ions, i.e., Co atoms, are shown, using a $2\times{1}\times{1}$ supercell for the AFM // $ab$ and AFM 1$\bm{Q}$ structures, and a $2\times{2}\times{1}$ supercell for the AFM AIAO structure.
As noted in Sec.~\ref{subsec:CoTS} of the main text, the AFM 1$\boldsymbol{Q}$ and AFM AIAO structures degenerate in the classical spin model only with the BL interactions.

We show the SDFT$+U$ calculation results in Figs. \ref{fig:CoTS-sdft-u} for (b) the energy per Co atom and (c) the magnetic moment length of Co atoms, respectively.
The AIAO structure is confirmed to be the most stable configuration, with the AFM 1$\bm{Q}$ structure having energies closer to those of the AIAO structure than other magnetic configurations.
While the spin-orbit coupling is omitted in our calculations, this is also confirmed even when the spin-orbit coupling is included~\cite{Takagi2023-uw}.
The magnitudes of the magnetic moment are nearly identical across all compared magnetic structures for a given $U$.
For the band structure calculations used to derive the tight-binding model, we set $U = 0.5$ eV to align the magnitude of the magnetic moment with the experimental value~\cite{Takagi2023-uw}.
How the value of $U$ affects the spin interactions ($J_{ij}$ and $B_{ij}$) is discussed in Appendix \ref{app:CoTS-jij-u}.

% ----------------------------------------------------------------------
\section{Effect of the $U$ Value on the Spin Interactions in \CoTS}\label{app:CoTS-jij-u}
% ----------------------------------------------------------------------

\begin{figure}[htbp]
    \centering
    \includegraphics[width=0.7\columnwidth, keepaspectratio]{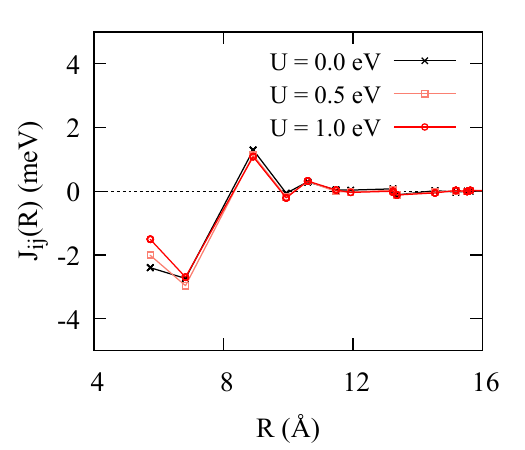}
    \caption{Distance dependence of the BL interactions for \CoTS at $\mu_{\text{DLM}} = 0$, showing how $J_{ij}$ changes when varying $U$ used in the \textit{ab initio} band structure calculations from which the tight-binding model is constructed.}
    \label{fig:CoTS-jij-r-u}
\end{figure}

\begin{figure}[htbp]
    \centering
    \includegraphics[width=\columnwidth, keepaspectratio]{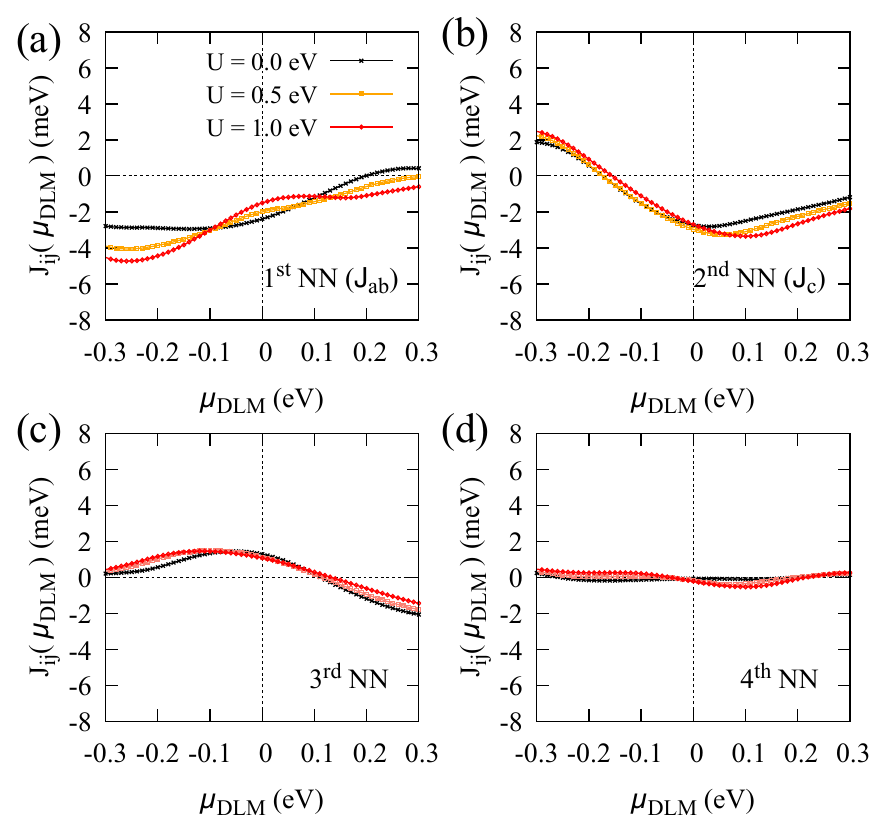}
    \caption{(a)-(d) Chemical potential dependence of BL interactions up to the fourth NNs for \CoTS, respectively.}
    \label{fig:CoTS-jij-mu-u}
\end{figure}

We here examine how the value of $U$ affects the calculated spin interactions by comparing the results for $U = 0$ and $U = 1$ eV.
In Figs. \ref{fig:CoTS-jij-r-u} and \ref{fig:CoTS-jij-mu-u}, we show the distance dependence of the BL interactions $(J_{ij}(R))$ at the chemical potential of the DLM state $(\muDLM=0)$ and the chemical potential dependence of those $(J_{ij}(\muDLM))$ up to the fourth NN interactions.
Both of the absolute values ($\sim{O}(1)$ meV) and the signs of the interactions do not change significantly with varying $U$.
Additionally, the trends with respect to tuning the chemical potential remain common across all values of $U$.
This demonstrates that the frustration on the HCP lattice, induced by antiferromagnetic BL interactions on the first and second NN bonds, is a robust feature for \CoTS.

\begin{figure}[htbp]
    \centering
    \includegraphics[width=0.95\columnwidth, keepaspectratio]{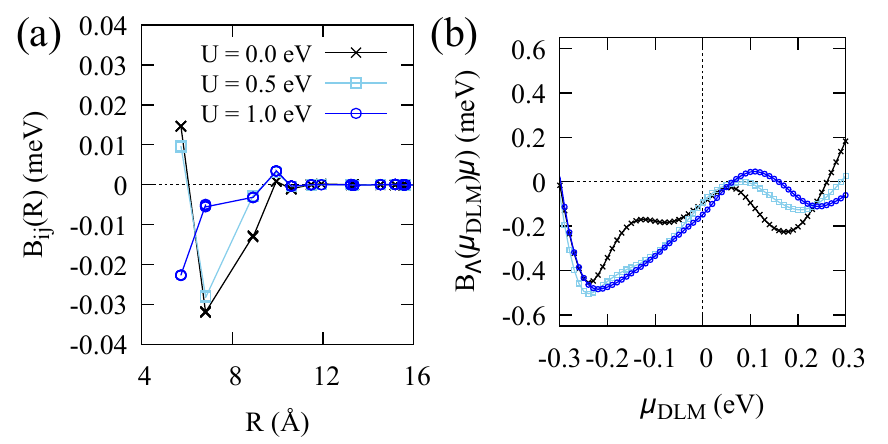}
    \caption{(a) Distance dependence of the BQ interaction at $\mu_{\text{DLM}} = 0$. (b) Chemical potential dependence of $B_{\Lambda}$ for \CoTS.}
    \label{fig:CoTS-comp-u-bij}
\end{figure}

\begin{figure}[htbp]
    \centering
    \includegraphics[width=\columnwidth, keepaspectratio]{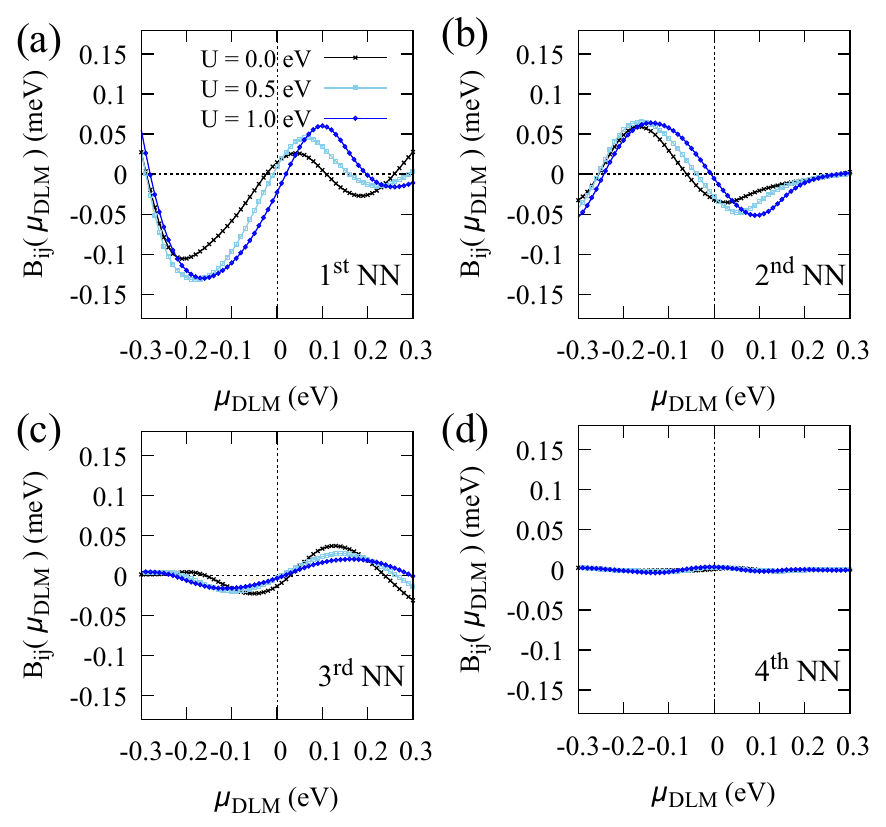}
    \caption{(a)-(d) Chemical potential dependence of BQ interactions for \CoTS up to the fourth NNs, respectively.}
    \label{fig:CoTS-comp-u-bij-mu}
\end{figure}

Next, we present the results for the BQ interactions in Fig. \ref{fig:CoTS-comp-u-bij}.
Notably, each individual BQ interaction depends on the $U$ value, both in terms of magnitude ($\sim O(10^{-2})$ meV) and sign, as shown in Figs. \ref{fig:CoTS-comp-u-bij} (a) \textr{and \ref{fig:CoTS-comp-u-bij-mu}}.
However, as presented in Fig. \ref{fig:CoTS-comp-u-bij} (b), $B_{\Lambda}$ ($\sim O(10^{-1})$ meV) remains relatively stable despite variations in $U$.
Specifically, the trend that $B_{\Lambda}$ tends to become positive with electron doping persists for all $U$ values.

In Fig. \ref{fig:CoTS-jij-mu-u}, we present the chemical potential dependence of each BQ interaction up to the fourth NNs.
The overall trend with respect to changes in the chemical potential is common for all BQ interactions.
However, due to the very small energy scale of the BQ interactions, both the magnitudes and the signs of the BQ interactions at $\muDLM=0$ change with varying $U$.

% ----------------------------------------------------------------------
\section{Effect of the Starting Magnetic Order}\label{app:comp-afm}
% ----------------------------------------------------------------------

As mentioned and explored for elemental magnetic metals in Sec.~\ref{sec:benchmark} in the main text, results of our method, i.e., SCE-DLM method based on \textit{ab initio} tight-binding model, could depend on the starting magnetic order from which the tight-binding model is constructed.
Therefore, we here show the results of our method for compounds in Sec.~\ref{sec:compounds} in the main text when we start from another magnetic order which is different from the one shown in the main text.

% ----------------------------------------------------------------------
\subsection{\CoTS}\label{subapp:CoTS-afm}

\begin{figure}[htbp]
    \centering
    \includegraphics[width=0.95\columnwidth, keepaspectratio]{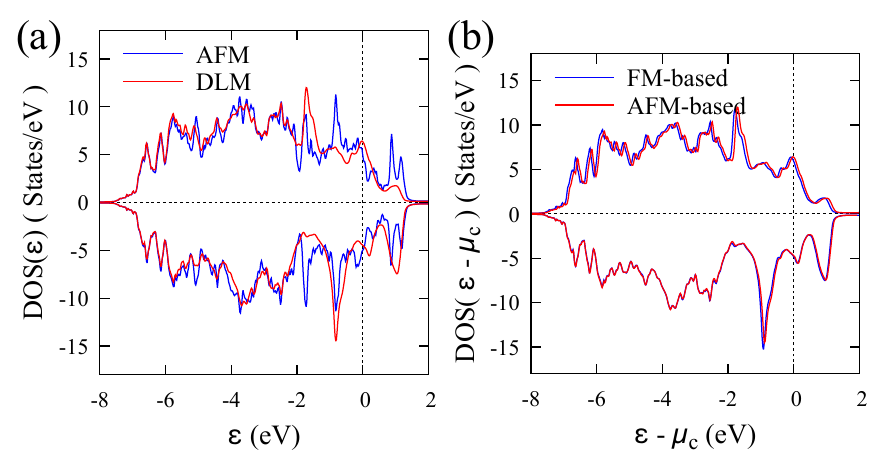}
    \caption{(a) DOS of the AFM state and the AFM-based DLM state. (b) DOS of the FM-based and AFM-based DLM states for \CoTS.}
    \label{fig:CoTS-comp-afm-dos}
\end{figure}

To examine the effect of the starting magnetic order on the calculated spin interactions, we compare results obtained from the FM state (shown in the main text) and the AFM // $c$ state, shown in Fig. \ref{fig:CoTS-sdft-u} (a).
We refer to the DLM states derived from these magnetic orders as FM-based and AFM-based DLM states, respectively.
First, Fig. \ref{fig:CoTS-comp-afm-dos} presents (a) the DOS of the AFM state and the AFM-based DLM state and (b) the DOS of the FM-based and AFM-based DLM states for comparison.
The DOS of the two DLM states, shown in Fig. \ref{fig:CoTS-comp-afm-dos} (b), are nearly identical.
The magnetic moments of Co atoms are 1.57 $\muB$ in the AFM state and 1.43 $\muB$ in the corresponding DLM state, which is nearly identical to the FM-based DLM state.
Therefore, the AFM-based DLM state is almost the same as the FM-based DLM state.

\begin{figure*}[htbp]
    \centering
    \includegraphics[width=1.9\columnwidth, keepaspectratio]{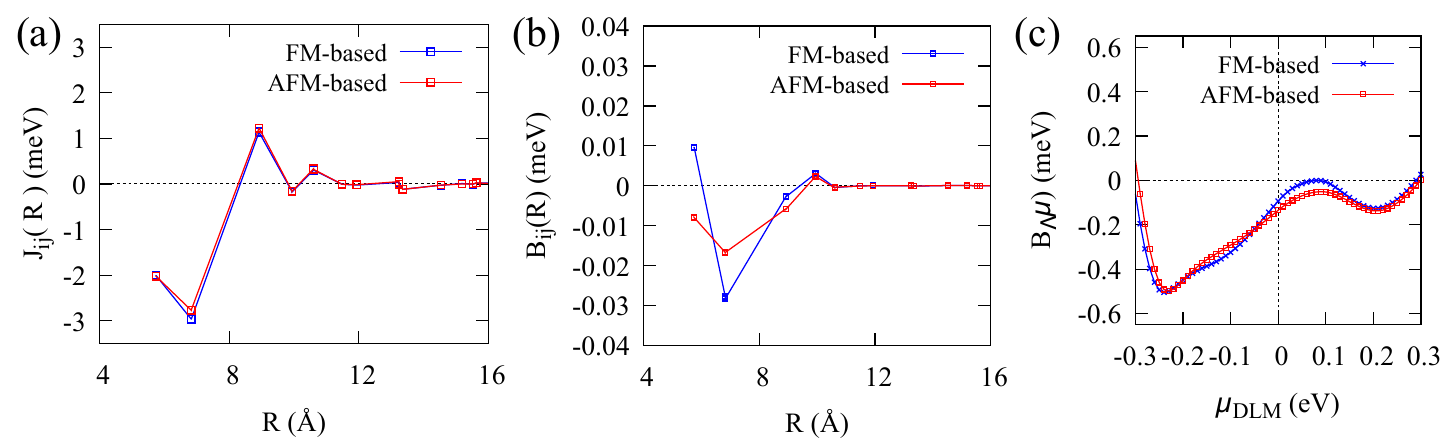}
    \caption{(a) Distance dependence of the BL interactions for \CoTS calculated from the FM-based and AFM-based DLM states. (b) Distance dependence of the BQ interactions calculated from the FM-based and AFM-based DLM states. (c) Chemical potential dependence of $B_{\Lambda}$ calculated from these two DLM states.}
    \label{fig:CoTS-comp-afm-jij-bij}
\end{figure*}

Then, in Fig. \ref{fig:CoTS-comp-afm-jij-bij}, we present the distance dependence of the (a) BL and (b) BQ interaction calculated from the two DLM states.
Fig. \ref{fig:CoTS-comp-afm-jij-bij} (c) shows the chemical potential dependence of the sum of the BQ interactions.
The BL interactions obtained from the two DLM states are found to be practically identical.
Unlike the BL interactions, the two DLM states yield different values for each BQ interaction.
However, the trend of the sum of the BQ interactions remains consistent against changes in the chemical potential.
From these results, our arguments in the main text based on the FM-based DLM state would remain valid, even when starting from other magnetically ordered states.

% ----------------------------------------------------------------------
\subsection{Pottasium Electrosodalite}\label{subapp:PES-afm}

\begin{figure}[htbp]
    \centering
    \includegraphics[width=0.95\columnwidth, keepaspectratio]{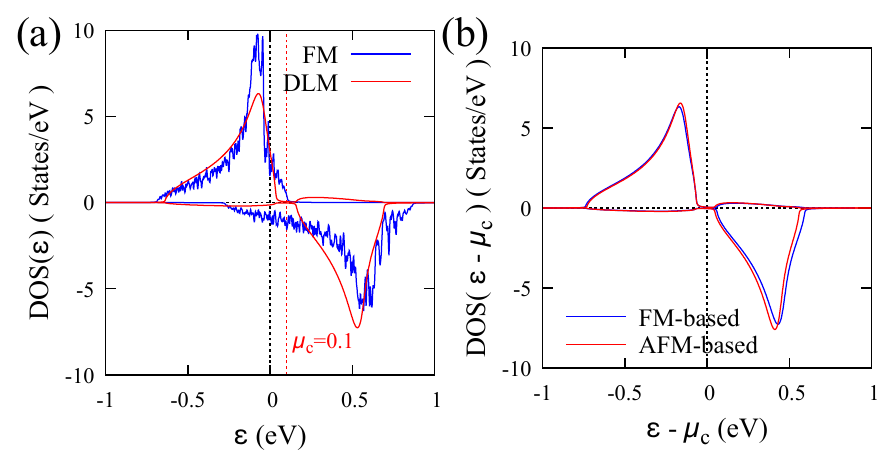}
    \caption{(a) DOS of the FM state and the FM-based DLM state. (b) DOS of the FM-based and AFM-based DLM states for \PES.}
    \label{fig:PES-comp-afm-dos}
\end{figure}

For \PES, we started from the AFM state in the main text.  
To investigate the effect of the initial magnetic configuration, we compare the results obtained from the AFM state with those from the FM state.  
Figure~\ref{fig:PES-comp-afm-dos} presents (a) the DOS of the FM state and the FM-based DLM state, and (b) a comparison of the FM-based and AFM-based DLM states.  
While the FM state itself is metallic, the FM-based DLM state is insulating, similar to the AFM-based DLM state.  
As shown in Fig.~\ref{fig:PES-comp-afm-dos}~(b), the FM-based and AFM-based DLM states exhibit nearly identical DOS.  
In both cases, the DLM state is insulating, and each interstitial $s$ orbital is well isolated by the surrounding aluminosilicate cage.  
As a result, the locally polarized magnetic moments are well-defined in this compound, and the DLM state is largely insensitive to the choice of the initial magnetic order.

\begin{figure}[htbp]
    \centering
    \includegraphics[width=0.95\columnwidth, keepaspectratio]{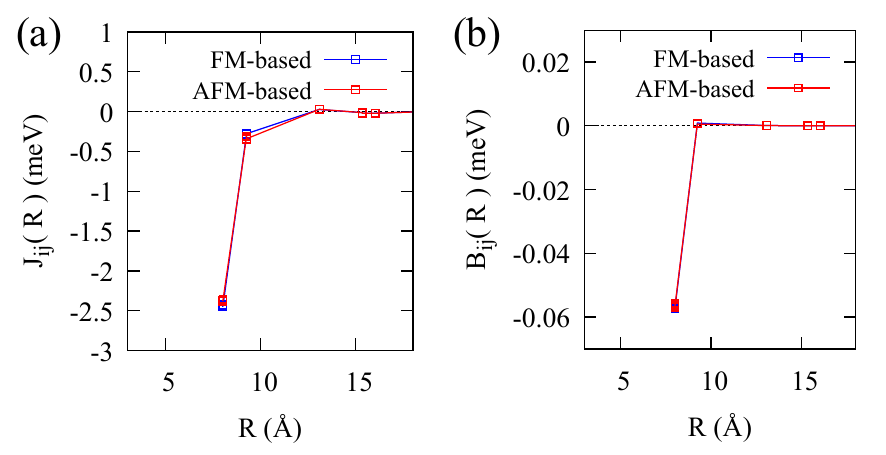}
    \caption{Distance dependence of the (a) BL and (b) BQ interactions calculated from the FM-based and AFM-based DLM states for \PES.}
    \label{fig:PES-comp-afm-jij-bij}
\end{figure}

Figure~\ref{fig:PES-comp-afm-jij-bij} shows the (a) BL and (b) BQ interactions obtained from the FM-based and AFM-based DLM states.  
Consistent with the DOS results, the spin interactions derived from both magnetic configurations are effectively indistinguishable.  
This further confirms that the starting magnetic order has minimal influence on the resulting spin interactions in this compound.

\bibliography{main}

\end{document}